\documentclass{pnastwo}
\usepackage{amsmath}
\usepackage{graphicx}
\newcommand{\onepic}[1]{\includegraphics[scale=0.24]{#1}}
\begin{document}
\title{Solving the apparent diversity-accuracy dilemma of
recommender systems}

\author{Tao Zhou\affil{1}{Department of Physics, University of
Fribourg, Chemin du Mus\'ee 3, CH-1700 Fribourg, Switzerland}
\affil{2}{Department of Modern Physics and Nonlinear Science
Center, University of Science and Technology of China, Hefei
230026, China}
\affil{3}{Research Center for Complex Systems Science,
University of Shanghai for Science and Technology, Shanghai
200093, China},
Zolt\'an Kuscsik\affil{1}{}\affil{4}{Department of
Theoretical Physics and Astrophysics, P.~J.~\v Saf\'arik
University, Park Angelinum 9, Ko\v sice 04001, Slovak Republic},
Jian-Guo Liu\affil{1}{}\affil{2}{}\affil{3}{},
Mat\'u\v s Medo\affil{1}{},
Joseph R.\ Wakeling\affil{1}{}\and
Yi-Cheng Zhang}

\contributor{Submitted to Proceedings of the National Academy
of Sciences of the United States of America}

\maketitle

\begin{article}

\begin{abstract}
Recommender systems use data on past user preferences to predict possible
future likes and interests.  A key challenge is that while the most useful
individual recommendations are to be found among diverse niche objects, the
most reliably accurate results are obtained by methods that recommend objects
based on user or object similarity.  In this paper we introduce a new algorithm
specifically to address the challenge of diversity and show how it can be used
to resolve this apparent dilemma when combined in an elegant hybrid with an
accuracy-focused algorithm.  By tuning the hybrid appropriately we are able to
obtain, without relying on any semantic or context-specific information,
simultaneous gains in both accuracy and diversity of recommendations.
\end{abstract}

\keywords{information filtering | recommender systems | hybrid
algorithms}


\dropcap{G}etting what you want, as the saying goes, is easy: the hard part is
working out what it is that you want in the first
place~\cite{Masum2004}.  Whereas information filtering tools like search
engines typically require the user to specify in advance what they are
looking for~\cite{Uri2001,Brin1998,Kleinberg1999,Belkin2000}, this challenge of
\emph{identifying} user needs is the domain of \emph{recommender
systems}~\cite{Belkin2000,Goldberg1992,Resnick1997,Adomavicius2005}, which
attempt to anticipate future likes and interests by mining data on past user
activities.

Many diverse recommendation techniques have been developed, including
collaborative filtering~\cite{Goldberg1992,Schafer2007}, content-based
analysis~\cite{Pazzani2007}, spectral analysis~\cite{Goldberg2001,Maslov2001},
latent semantic models and Dirichlet allocation~\cite{Hofmann2004,Blei2003},
and iterative self-consistent
refinement~\cite{Laureti06,deKerchove2008,Ren2008}.  What most have in common
is that they are based on \emph{similarity}, either of users or objects or
both: for example, e-commerce sites such as Amazon.com use the overlap between
customers' past purchases and browsing activity to recommend
products~\cite{Schafer2001,Linden2003}, while the TiVo digital video system
recommends TV shows and movies on the basis of correlations in users' viewing
patterns and ratings~\cite{Ali2004}.  The risk of such an approach is that,
with recommendations based on overlap rather than difference, more and more
users will be exposed to a narrowing band of popular objects, while niche items
that might be very relevant will be overlooked.

The focus on similarity is compounded by the metrics used to assess
recommendation performance.  A typical method of comparison is to consider an
algorithm's accuracy in reproducing known user opinions that have been removed
from a test data set.  An \emph{accurate} recommendation, however, is not
necessarily a \emph{useful} one: real value is found in the ability to suggest
objects users would not readily discover for themselves, that is, in the
novelty and diversity of recommendation~\cite{Herlocker2004}.  Despite this,
most studies of recommender systems focus overwhelmingly on accuracy as the
only important factor (for example, the Netflix Prize~\cite{Bennett2007}
challenged researchers to increase accuracy without any reference to novelty or
personalization of results).  Where diversification is addressed, it is
typically as an adjunct to the main recommendation process, based on
restrictive features such as semantic or other context-specific
information~\cite{Burke2002,Ziegler2005}.

The clear concern is that an algorithm that focuses too strongly on diversity
rather than similarity is putting accuracy at risk.  Our main focus in this
paper is to show that this apparent dilemma can in fact be resolved by an
appropriate combination of accuracy- and diversity-focused methods.  We begin
by introducing a ``heat-spreading'' algorithm designed specifically to address
the challenge of diversity, with high success both at seeking out novel items
and at enhancing the personalization of individual user recommendations.  We
show how this algorithm can be coupled in a highly efficient hybrid with a
diffusion-based recommendation method recently introduced by our
group~\cite{Zhou2007}.  Using three different datasets from three distinct
communities, we employ a combination of accuracy- and diversity-related metrics
to perform a detailed study of recommendation performance and a comparison to
well-known methods.  We show that not only does the hybrid algorithm outperform
other methods but that, without relying on any semantic or context-specific
information, it can be tuned to obtain significant and simultaneous gains in
both accuracy and diversity of recommendations.

\section{Methods}
\subsection{Recommendation procedure}
Since explicit ratings are not always available~\cite{Claypool2001}, the
algorithms studied in this paper are selected to work with very simple input
data: $u$ users, $o$ objects, and a set of links between the two corresponding
to the objects collected by particular users (more explicit preference
indicators can be easily mapped to this ``unary'' form, albeit losing
information in the process, whereas the converse is not so). These links can be
represented by an $o\times u$ adjacency matrix $\mathsf{A}$ where
$a_{\alpha i}=1$ if object $\alpha$ is collected by user $i$ and
$a_{\alpha i}=0$ otherwise (throughout this paper we use Greek and Latin
letters respectively for object- and user-related indices). Alternatively we
can visualize the data as a bipartite user-object network with $u+o$ nodes,
where the degrees of object and user nodes, $k_{\alpha}$ and $k_{i}$, represent
respectively the number of users who have collected object $\alpha$ and the
number of objects collected by user $i$.

Recommendation scores are calculated for each user and each of
their uncollected objects, enabling the construction of a sorted
recommendation list with the most-recommended items at the top.
Different algorithms generate different object scores and thus
different rankings.

\subsection{Algorithms}
The \emph{heat spreading} (HeatS) algorithm introduced here employs
a process analogous to heat diffusion across the user-object network.
This can be related to earlier work using a ``heat conduction'' algorithm
to generate recommendations~\cite{Blattner07,Stojmirovic07}, but with some
key differences.  The earlier algorithm operates on an object-object network
derived from an explicit ratings structure, which washes out information about
novelty or popularity of objects and consequently limits the algorithm to
considering questions of accuracy and not diversity.  The algorithm also
requires multiple iterations to converge to a steady state.  By contrast HeatS
requires no more than unary data, and generates effective recommendations in a
single pass.

HeatS works by assigning objects an initial level of ``resource''
denoted by the vector $\boldsymbol{f}$ (where $f_{\beta}$ is the
resource possessed by object $\beta$), and then redistributing it
via the transformation
$\tilde{\boldsymbol{f}}=\mathsf{W}^H\boldsymbol{f}$, where
\begin{equation}
\label{WH}
W_{\alpha\beta}^H=\frac1{k_{\alpha}}\sum_{j=1}^u
\frac{a_{\alpha j}a_{\beta j}}{k_j}
\end{equation}
is a row-normalized $o\times o$ matrix representing a discrete
analogy of a heat diffusion process.  Recommendations for a given
user $i$ are obtained by setting the initial resource vector
$\boldsymbol{f}^{i}$ in accordance with the objects the user has
already collected, that is, by setting
$f_{\beta}^{i} = a_{\beta i}$. The resulting recommendation list
of uncollected objects is then sorted according to
$\tilde{f}_{\alpha}^{i}$ in descending order.

HeatS is a variant on an earlier \emph{probabilistic spreading}
(ProbS) algorithm introduced by our group~\cite{Zhou2007}, which
redistributes resource in a manner akin to a random walk process.
Whereas HeatS employs a row-normalized transition matrix, that
of ProbS is column-normalized,
\begin{equation}
\label{WP}
W_{\alpha\beta}^P=\frac1{k_{\beta}}\sum_{j=1}^u
\frac{a_{\alpha j}a_{\beta j}}{k_j},
\end{equation}
with the resource redistribution and resulting object scores then
being given by $\tilde{\boldsymbol{f}}=\mathsf{W}^P\boldsymbol{f}
= (\mathsf{W}^H)^T\boldsymbol{f}$.

\begin{figure}
\centering
\includegraphics[width=0.45\textwidth]{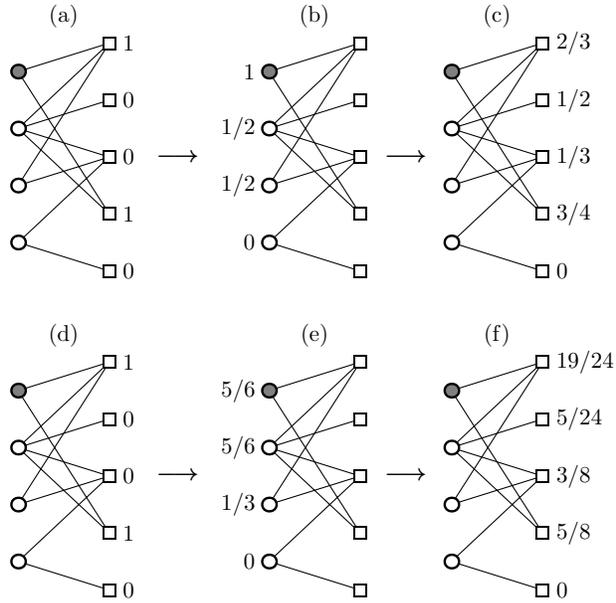}
\caption{The HeatS (a,b,c) and ProbS (d,e,f) algorithms
(Eqs.\ \ref{WH} and \ref{WP}) at work on the bipartite
user-object network.  Objects are shown as squares, users as
circles, with the target user indicated by the shaded circle.
While the HeatS algorithm redistributes resource via a
nearest-neighbour averaging process, the ProbS algorithm
works by an equal distribution of resource among nearest
neighbours.}
\label{fig:diffusion}
\end{figure}

A visual representation of the resource spreading processes of
ProbS and HeatS is given in Fig.\ \ref{fig:diffusion}: in ProbS
(a--c) the initial resource placed on objects is first evenly
distributed among neighboring users, and then evenly
redistributed back to those users' neighboring objects.  By
contrast HeatS (d--f) redistributes resource via an averaging
procedure, with users receiving a level of resource equal to the
mean amount possessed by their neighboring objects, and objects
then receiving back the mean of their neighboring users'
resource levels.  (Note that in ProbS total resource levels
remain constant, whereas in HeatS this is not so.)
Due to the sparsity of real datasets, these ``physical''
descriptions of the algorithms turn out to be more
computationally efficient in practice than constructing and using
the transition matrices $\mathsf{W}^P$ and $\mathsf{W}^H$.

To provide a point of comparison we also employ two methods
well-known in the recommender systems literature.
\emph{Global ranking} (GRank) recommends objects according to
their overall popularity, sorting them by their degree $k_{\alpha}$
in descending order.  While computationally cheap, GRank is not
personalized (apart from the exclusion of different already-collected
objects) and in most cases it performs poorly.

A much more effective method is \emph{user similarity} (USim), a well
known and widely used technique that recommends items frequently
collected by a given user's ``taste mates''~\cite{Adomavicius2005}.
The taste overlap between users $i$ and $j$ is measured by the cosine
similarity,
\begin{equation}
\label{uusimilarity}
s_{ij}=\frac{\sum_{\alpha=1}^o a_{\alpha i}a_{\alpha j}}
{\sqrt{k_i k_j}},
\end{equation}
and if user $i$ has not yet collected object $\alpha$, its
recommendation score is given by
\begin{equation}
\label{USim}
v_{\alpha i}=\frac{\sum_{j=1}^u s_{ij}a_{\alpha j}}
{\sum_{j=1}^u s_{ij}},
\end{equation}
with the final recommendation list for user $i$ being sorted
according to $v_{\alpha i}$ in descending order.

\subsection{Hybrid methods}
A basic but very general means of creating hybrid algorithms is to use weighted
linear aggregation~\cite{Burke2002}: if methods X and Y report scores of
$x_{\alpha}$ and $y_{\alpha}$ respectively, then a hybrid score for object
$\alpha$ can be given by
\begin{equation}
\label{eq:hybrid}
z_{\alpha} =
(1-\lambda)\left[\frac{x_{\alpha}}{\max_{\beta}x_{\beta}}
\right]
+
\lambda\left[\frac{y_{\alpha}}{\max_{\beta}y_{\beta}}\right],
\end{equation}
where the normalizations address the fact that different methods may produce
scores on very different scales.  By varying the parameter $\lambda \in [0,1]$,
we can tune the hybrid X+Y to favor the characteristics of one method or the
other.

Though easy to implement, this approach has the disadvantage of
requiring two independent recommendation calculations, thus
increasing computational cost.  HeatS and ProbS, however, are
already fundamentally linked, with their recommendation processes
being determined by different normalizations of the same underlying
matrix (in fact, their transition matrices are the transpose of
each other).  A much more elegant hybrid can thus be achieved by
incorporating the hybridization parameter $\lambda$ into the
transition matrix normalization:
\begin{equation}
\label{Whybrid1}
W_{\alpha\beta}^{H+P} =
\frac1{k_{\alpha}^{1-\lambda}k_{\beta}^{\lambda}}
\sum_{j=1}^u \frac{a_{\alpha j}a_{\beta j}}{k_j},
\end{equation}
where $\lambda=0$ gives us the pure HeatS algorithm, and
$\lambda=1$ gives us pure ProbS (other hybrid forms are
possible but give inferior performance: Fig.\ S1 of supporting
information [SI] provides a comparison of the different
alternatives). In contrast to Eq.\ \ref{eq:hybrid}, this
HeatS+ProbS hybrid has a computational complexity of order no
greater than ProbS or HeatS alone.  Note that while in the
present work $\lambda$ takes a universal value, there is no
reason in principle why we cannot use different values for each
individual target user.

\subsection{Datasets}
Three different datasets (Table~\ref{tab:datasets}) were used to
test the above algorithms, differing both in subject matter
(movies, music and internet bookmarks) and in quantitative
aspects such as user/object ratios and link sparsity.  The first
(\emph{Netflix}) is a randomly-selected subset of the huge
dataset provided for the Netflix Prize~\cite{Bennett2007}, while
the other two (\emph{RYM} and \emph{Delicious}) were obtained by
downloading publicly-available data from the music ratings
website RateYourMusic.com and the social bookmarking website
Delicious.com (taking care to anonymize user identity in the
process).

While the Delicious data is inherently unary (a user has either
collected a web link or not), the raw Netflix and RYM data
contain explicit ratings on a 5-star scale.  A coarse-graining
procedure was therefore used to transform these into unary form:
an object is considered to be collected by a user only if the
given rating is 3 or more.  Sparseness of the datasets (defined
as the number of links divided by the total number of possible
user-object pairs) is measured relative to these coarse-grained
connections.

\subsection{Recommendation performance metrics}
To test a recommendation method on a dataset we remove at random
10\%\ of the links and apply the algorithm to the remainder
to produce a recommendation list for each user.  We then employ
four different metrics, two to measure accuracy in recovery of
deleted links (A) and two to measure recommendation diversity
(D):

(A1) \emph{Recovery of deleted links}, $r$.  An accurate method
will clearly rank preferable objects more highly than disliked
ones.  Assuming that users' collected objects are indeed
preferred, deleted links should be ranked higher on average than
the other uncollected objects.  So, if uncollected object
$\alpha$ is listed in place $p$ for user $i$, the relative
rank $r_{\alpha i} = p/(o - k_{i})$ should be smaller if $\alpha$
is a deleted link (where objects from places $p_1$ to $p_2$ have
the same score, which happens often in practice, we give them all
the same relative ranking, $\frac{1}{2}[p_1 + p_2]/[o-k_{i}]$).
Averaging over all deleted links we obtain a quantity, $r$, such
that the smaller its value, the higher the method's ability to
recover deleted links.

(A2) \emph{Precision and recall enhancement}, $e_{P}(L)$ and
$e_{R}(L)$. Since real users usually consider only the top part
of the recommendation list, a more practical measure may be to
consider $d_{i}(L)$, the number of user $i$'s deleted links
contained in the top $L$ places.  Depending on our concerns, we
may be interested either in how many of these top $L$ places are
occupied by deleted links, or how many of the user's $D_i$
deleted links have been recovered in this way.  Averaging these
ratios $d_{i}(L)/L$ and $d_{i}(L)/D_i$ over all users with at
least one deleted link, we obtain the mean \emph{precision} and
\emph{recall}, $P(L)$ and $R(L)$, of the recommendation
process~\cite{Herlocker2004,Swets1963}.

A still better perspective may be given by considering these
values relative to the precision and recall of random
recommendations, $P_{\mathrm{rand}}(L)$ and
$R_{\mathrm{rand}}(L)$. If user $i$ has a total of $D_{i}$
deleted links, then $P_{\mathrm{rand}}^{i}(L) = D_{i}/(o - k_{i})
\approx D_{i}/o$ (since in general $o \gg k_{i}$) and hence
averaging over all users, $P_{\mathrm{rand}}(L) = D/(ou)$, where
$D$ is the total number of deleted links.  By contrast the
mean number of deleted links in the top $L$ places is given by
$L D_{i}/(o-k_{i}) \approx L D_{i}/o$ and so
$R_{\mathrm{rand}}(L) = L/o$.  From this we can define the
precision and recall \emph{enhancement},
\begin{subequations}
\begin{eqnarray}
e_{P}(L) & := & \frac{P(L)}{P_{\mathrm{rand}}(L)} = P(L) \cdot
\frac{ou}{D}, \\
e_{R}(L) & := & \frac{R(L)}{R_{\mathrm{rand}}(L)} = R(L) \cdot
\frac{o}{L}.
\end{eqnarray}
\end{subequations}

Results for recall are given in SI (Figs.\ S2 and S3), but are
similar in character to those shown here for precision.

(D1) \emph{Personalization}, $h(L)$.  Our first measure of
diversity considers the uniqueness of different users'
recommendation lists---that is, inter-user diversity. Given two
users $i$ and $j$, the difference between their recommendation
lists can be measured by the inter-list distance,
\begin{equation}
h_{ij}(L) := 1 - \frac{q_{ij}(L)}{L},
\end{equation}
where $q_{ij}(L)$ is the number of common items in the top $L$
places of both lists: identical lists thus have $h_{ij}(L) = 0$
whereas completely different lists have $h_{ij}(L) = 1$.
Averaging $h_{ij}(L)$ over all pairs of users with at least
one deleted link we obtain the mean distance $h(L)$, for which
greater or lesser values mean respectively greater or lesser
personalization of users' recommendation lists.

(D2) \emph{Surprisal/novelty}, $I(L)$.  The second type of diversity concerns
the capacity of the recommender system to generate novel and unexpected
results---to suggest objects a user is unlikely to already know about.  To
measure this we use the self-information or ``surprisal''~\cite{Tribus1961} of
recommended objects, which measures the unexpectedness of an object relative to
its global popularity. Given an object $\alpha$, the chance a randomly-selected
user has collected it is given by $k_{\alpha}/u$ and thus its self-information
is $I_{\alpha} = \log_{2} (u/k_{\alpha})$.  From this we can calculate the mean
self-information $I_{i}(L)$ of each user's top $L$ objects, and averaging over
all users with at least one deleted link we obtain the mean top-$L$ surprisal
$I(L)$.

Note that unlike the metrics for accuracy, the diversity-related
measures could be averaged over all users regardless of whether
they have deleted links or not, but the final results do not
differ significantly. Where metrics depend on $L$, different
choices result in shifts in the precise numbers but relative
performance differences between methods remain unchanged so long
as $L \ll o$. Extended results are available in SI (Figs.\ S4
and S5); a value of $L=20$ was chosen for the results
displayed here in order to reflect the likely length of a
practical recommendation list.

\section{Results}

\subsection{Individual algorithms}
A summary of the principal results for all algorithms, metrics
and datasets is given in Table~\ref{tab:comparison}.

ProbS is consistently the strongest performer with respect to
accuracy, with USim a close second, while both GRank and HeatS
perform significantly worse (the latter reporting particularly
bad performance with respect to precision enhancement).  By
contrast with respect to the diversity metrics HeatS is by far
the strongest performer: ProbS has some success with respect to
personalization, but along with USim and GRank performs weakly
where surprisal (novelty) is concerned.

That GRank has any personalization at all ($h(L)>0$) stems only
from the fact that it does not recommend items already collected,
and different users have collected different items.  The
difference in GRank's performance between Netflix, RYM and
Delicious can be ascribed to the ``blockbuster'' phenomenon
common in movies, far less so with music and web links: the 20
most popular objects in Netflix are each collected by on average
31.7\%\ of users, while for RYM the figure is 7.2\%\ and for
Delicious only 5.6\%.

The opposing performances of ProbS and HeatS---the former
favoring accuracy, the latter personalization and novelty---can
be related to their different treatment of popular objects.  The
random-walk procedure of ProbS favors highly-connected objects,
whereas the averaging process of HeatS favors objects with few
links: for example, in the Delicious dataset the average degree
of users' top 20 objects as returned by ProbS is 346, while with
HeatS it is only 2.2. Obviously the latter will result in high
surprisal values, and also greater personalization, as low-degree
objects are more numerous and a method that favors them has a
better chance of producing different recommendation lists for
different users. On the other hand randomly-deleted links are
clearly more likely to point to popular objects, and methods that
favor low-degree objects will therefore do worse; hence the
indiscriminate but populist GRank is able to outperform the
novelty-favoring HeatS.

If we deliberately delete only links to low-degree objects, the
situation is reversed, with HeatS providing better accuracy,
although overall performance of all algorithms deteriorates
(Table~\ref{tab:low-degree} and Fig.\ S6). Hence, while populism
can be a cheap and easy way to get superficially accurate results,
it is limited in scope: the most appropriate method can be determined
only in the context of a given task or user need.  The result
also highlights the very distinct and unusual character of HeatS
compared to other recommendation methods.

\subsection{Hybrid methods}
\begin{figure*}
\centering
\includegraphics[width=0.9\textwidth]{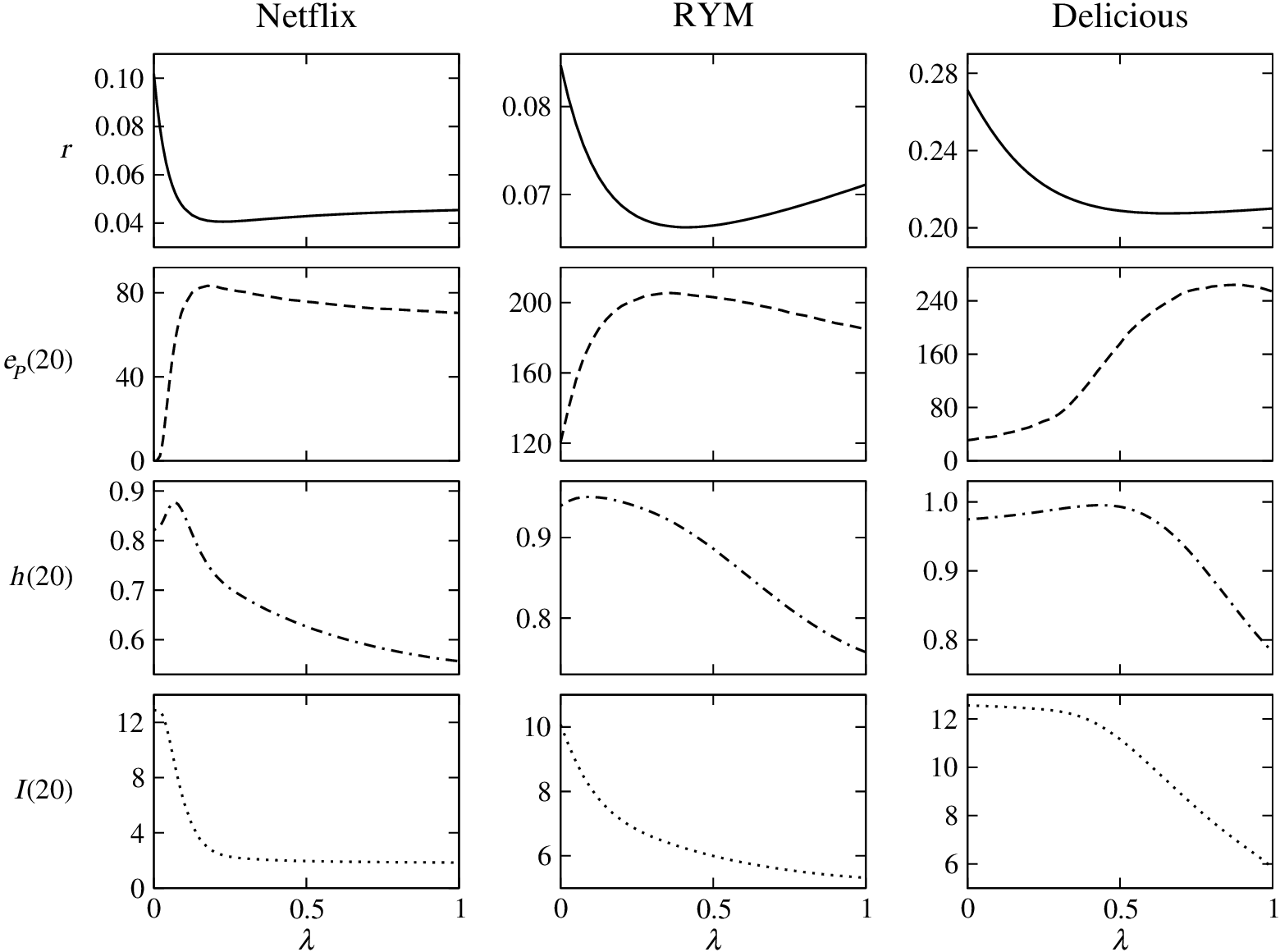}
\caption{Performance of the HeatS+ProbS hybrid algorithm
(Eq.\ \ref{Whybrid1}) on the three different datasets.  By
varying the hybridization parameter between pure HeatS
($\lambda=0$) and pure ProbS ($\lambda=1$) it is possible to gain
simultaneous performance enhancements with respect to both
accuracy ($r$ and $e_{P}(L)$) and diversity ($h(L)$ and $I(L)$)
of recommendations. Tuning $\lambda$ in this fashion allows the
algorithm to be customized and optimized for different user or
community needs.}
\label{fig:hybrid}
\end{figure*}

Given that different algorithms serve different purposes and
needs, is it possible to combine two (or more) in such a way as
to obtain the best features of both?  With HeatS favoring
diversity and ProbS accuracy, their hybrid combination
(Eq.\ \ref{Whybrid1}) might be expected to provide a smooth
transition from one to the other. In fact, the situation is even
more favorable: while pure HeatS represents the optimum for
novelty, it is possible to obtain performance improvements
relative to all other metrics by tuning the hybridization
parameter $\lambda$ appropriately (Fig.\ \ref{fig:hybrid}).  The
accuracy of ProbS can thus be maintained and even improved while
simultaneously attaining diversity close to or even exceeding
that of HeatS. Alternatively, diversity can be favored while
minimizing the cost in terms of accuracy.

Depending on the particular needs of a system and its users, one
can define an arbitrary utility function $U(r,e_{P},h,I,L)$ and
choose $\lambda$ to optimize it: {Table}~\ref{tab:improvements}
gives as an example the percentage improvements that can be made,
relative to pure ProbS ($\lambda=1$), if we choose $\lambda$ to
minimize $r$.  Shared improvements are obtained for all metrics
except with the Delicious dataset, where minimizing $r$ has a
negative effect on $e_{P}(L)$.  However, from
Fig.\ \ref{fig:hybrid} we can see that even in this case it is
possible to choose a value of $\lambda$ to simultaneously improve
all metrics relative to ProbS.

Although HeatS+ProbS provides the best performance when taking
into account all the metrics, other hybrids (constructed using
the more general method of Eq.\ \ref{eq:hybrid}) can provide
some valuable individual contributions (Fig.\ S7). HeatS+USim
behaves similarly to HeatS+ProbS, but with generally smaller
performance improvements. A more interesting hybrid is to
combine the poorly-performing GRank with either HeatS or ProbS.
These combinations can have a dramatic effect on link recovery:
for RYM either can be tuned to produce an improvement in $r$ of
almost 30\%\ (relative to pure ProbS), compared to only 6.8\%\
for the HeatS+ProbS hybrid (Table~\ref{tab:improvements}).

The explanation for these improvements stems from the way in
which ProbS and HeatS interact with sparse datasets.  Coverage of
uncollected objects is limited to those sharing a user in common
with an object collected by the target user
(Fig.\ \ref{fig:diffusion}): all others receive a score of zero
and so share a common (and large) relative rank,
$r_{\alpha i} = (o - \frac{1}{2}(Z-1))/(o - k_{i})$ where
$Z$ is the number of objects with zero score.  GRank, with its
universal coverage, is able to differentially rank these objects
and so lower their contributions to $r$.  Consequently, while
incorporating it too strongly has a deleterious effect on the
other metrics, a small GRank contribution can provide a useful
enhancement to recommendation coverage---notably in ``cold
start'' cases where little or nothing is known about a user.

\section{Discussion}
Recommender systems have at their heart some very simple and natural social
processes.  Each one of us looks to others for advice and opinions, learning
over time who to trust and whose suggestions to discount.  The paradox is that
many of the most valuable contributions come not from close friends but from
people with whom we have only a limited connection---``weak ties'' who alert us
to possibilities outside our regular experience~\cite{Granovetter1973}.

The technical challenges facing recommender systems involve
similar paradoxes. The most reliably accurate algorithms are
those based on similarity and popularity of users and objects,
yet the most valuable recommendations are those of niche items
users are unlikely to find for themselves~\cite{Herlocker2004}.
In this paper we have shown how this apparent dilemma can be
resolved by an appropriate combination of diversity- and
accuracy-focused methods, using a hybrid algorithm that joins a
method with proven high accuracy with a new algorithm dedicated
specifically to the production of novel and personalized
recommendations. Their combination allows not merely a
compromise between the two imperatives but allows us to
simultaneously increase both accuracy and diversity of
recommendations.  By tuning the degree of hybridization the
algorithms can be tailored to many custom situations and
requirements.

We expect these results to be general: while we have presented a particular set
of algorithms and datasets here, other recommender systems must face the same
apparent dilemma and we expect them to benefit from a similar hybrid approach.
It is interesting to note that while the Netflix Prize focused solely on
accuracy, the winning entry in fact took a diversification approach, in this
case based on tracking the changes in user opinions over
time~\cite{NetflixWinners}.

The algorithms presented here rely on no more than unary data and can thus
place diversity at the heart of the recommendation process while still being
applicable to virtually any dataset.  More detailed sources of information can
nevertheless be used to extend the recommendation process. Topical information
and other measures of item-item similarity can be used to further diversify
recommendation lists~\cite{Ziegler2005}: user-generated classifications such as tags~\cite{Hotho2006,Cattuto2007,Zhang2010} may be useful here.  The HeatS and
ProbS algorithms, and hence their hybrid, can be further customized by
modifying the initial allocation of resource~\cite{Zhou2008} to increase or
decrease the influence of selected objects on the recommendation process.  The
hybridization process itself can be extended by incorporating techniques such
as content-based or semantic analyses~\cite{Burke2002}.

The ultimate measure of success for any recommender system is of
course in the appreciation of its users, and in particular the
ability of the system to serve their often very distinct needs.
While in this paper we have optimized the hybrid from a global
perspective, there is no reason why it cannot be tuned
differently for each individual user---either by the system
provider or by users themselves.  This last consideration opens
the door to extensive future theoretical and empirical research,
bringing diversity and personalization not just to the contents
of recommendation lists, but to the recommendation process
itself.

\begin{acknowledgments}
We are grateful to Yi-Kuo Yu for useful comments and conversations, and to two
anonymous referees for their valuable feedback.
This work was supported by  Swiss National Science Foundation grant
200020-121848, Swiss State Ministry for Education and Research grant C05.0148
(Physics of Risk), and National Natural Science Foundation of China grants
10635040 and 60744003.  We also acknowledge funding from the Liquid
Publications and QLectives projects (EU FET-Open grants 213360 and 231200)
during the final stages of this work.
\end{acknowledgments}

\end{article}

\begin{table}
\caption{Properties of the tested datasets.}
\label{tab:datasets}
\begin{tabular}{lrrrr}
dataset          & users   & objects  & links        &
sparsity\\
\hline
Netflix   & 10\,000 & 6\,000   & 701\,947 &
$1.17\cdot10^{-2}$\\
RYM       & 33\,786 & 5\,381   & 613\,387 &
$3.37\cdot10^{-3}$\\
Delicious & 10\,000 & 232\,657 & 1\,233\,997 &
$5.30\cdot10^{-4}$\\
\hline
\end{tabular}
\end{table}

\begin{table*}
\caption{Performance of the recommendation algorithms according
to each of the four metrics: recovery of deleted links,
precision enhancement, personalization, and surprisal.}
\label{tab:comparison}
\begin{tabular*}{\hsize}{@{\extracolsep{\fill}}rcccccccccccc}
 & \multicolumn{4}{c}{Netflix} & \multicolumn{4}{c}{RYM}
 & \multicolumn{4}{c}{Delicious}\\
method & $r$ & $e_{P}(20)$ & $h(20)$ & $I(20)$
       & $r$ & $e_{P}(20)$ & $h(20)$ & $I(20)$
       & $r$ & $e_{P}(20)$ & $h(20)$ & $I(20)$\\
\hline
GRank & 0.057 & 58.7  & 0.450 & 1.79
      & 0.119 & 57.3  & 0.178 & 4.64
      & 0.314 & 147   & 0.097 & 4.23\\
USim  & 0.051 & 68.5  & 0.516 & 1.82
      & 0.087 & 150   & 0.721 & 5.17
      & 0.223 & 249   & 0.522 & 4.49\\
ProbS & 0.045 & 70.4  & 0.557 & 1.85
      & 0.071 & 185   & 0.758 & 5.32
      & 0.210 & 254   & 0.783 & 5.81\\
HeatS & 0.102 & 0.11  & 0.821 & 12.9
      & 0.085 & 121   & 0.939 & 10.1
      & 0.271 & 30.8  & 0.975 & 12.6\\
\hline
\end{tabular*}
\end{table*}

\begin{table}
\caption{Performance of individual recommendation algorithms
for a probe set consisting of only low-degree ($k<100$) objects.}
\label{tab:low-degree}
\begin{tabular}{rcccc}
method & $r$   &$e_P(20)$&$h(20)$&$I(20)$\\
\hline
 GRank & 0.327 & 0.000   & 0.525 & 1.68\\
 USim  & 0.308 & 0.000 & 0.579 & 1.72\\
 ProbS & 0.279 & 0.014   & 0.610 & 1.74\\
 HeatS & 0.262 & 0.679   & 0.848 & 13.1\\
\hline
\end{tabular}
\end{table}

\begin{table}
\caption{Tuning the HeatS+ProbS hybridization parameter
$\lambda$ to optimize for $r$ produces simultaneous improvements
in other metrics.  The relative changes are given in percentage
terms against the pure ProbS algorithm.}
\label{tab:improvements}
\begin{tabular}{lrrrrr}
dataset   & $\lambda$ & $\delta r$ & $\delta e_{P}(20)$ &
$\delta h(20)$ & $\delta I(20)$\\
\hline
Netflix   & 0.23 & 10.6\% & 16.5\% & 28.5\% & 28.8\%\\
RYM       & 0.41 &  6.8\% & 10.8\% & 20.1\% & 17.2\%\\
Delicious & 0.66 &  1.2\% & -6.0\% & 22.5\% & 61.7\%\\
\hline
\end{tabular}
\end{table}

\clearpage

\noindent{\sf\bfseries
{\LARGE Supporting information}\\[8pt]
{\Large Zhou et al., ``Solving the apparent diversity-accuracy
dilemma\\
of recommender systems''}}

\vfill

\begin{center}
\begin{tabular}{rr}
\onepic{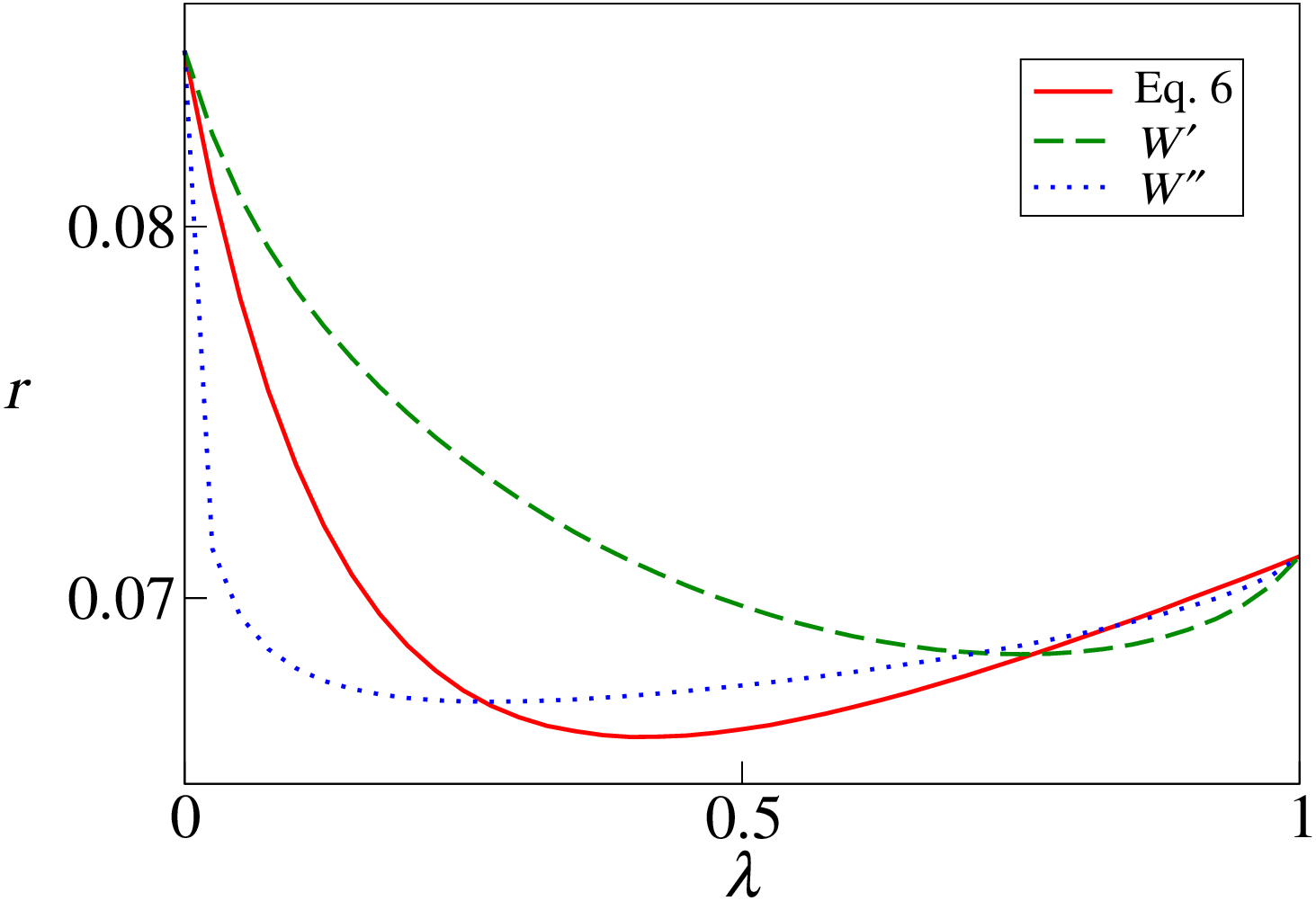} &
\onepic{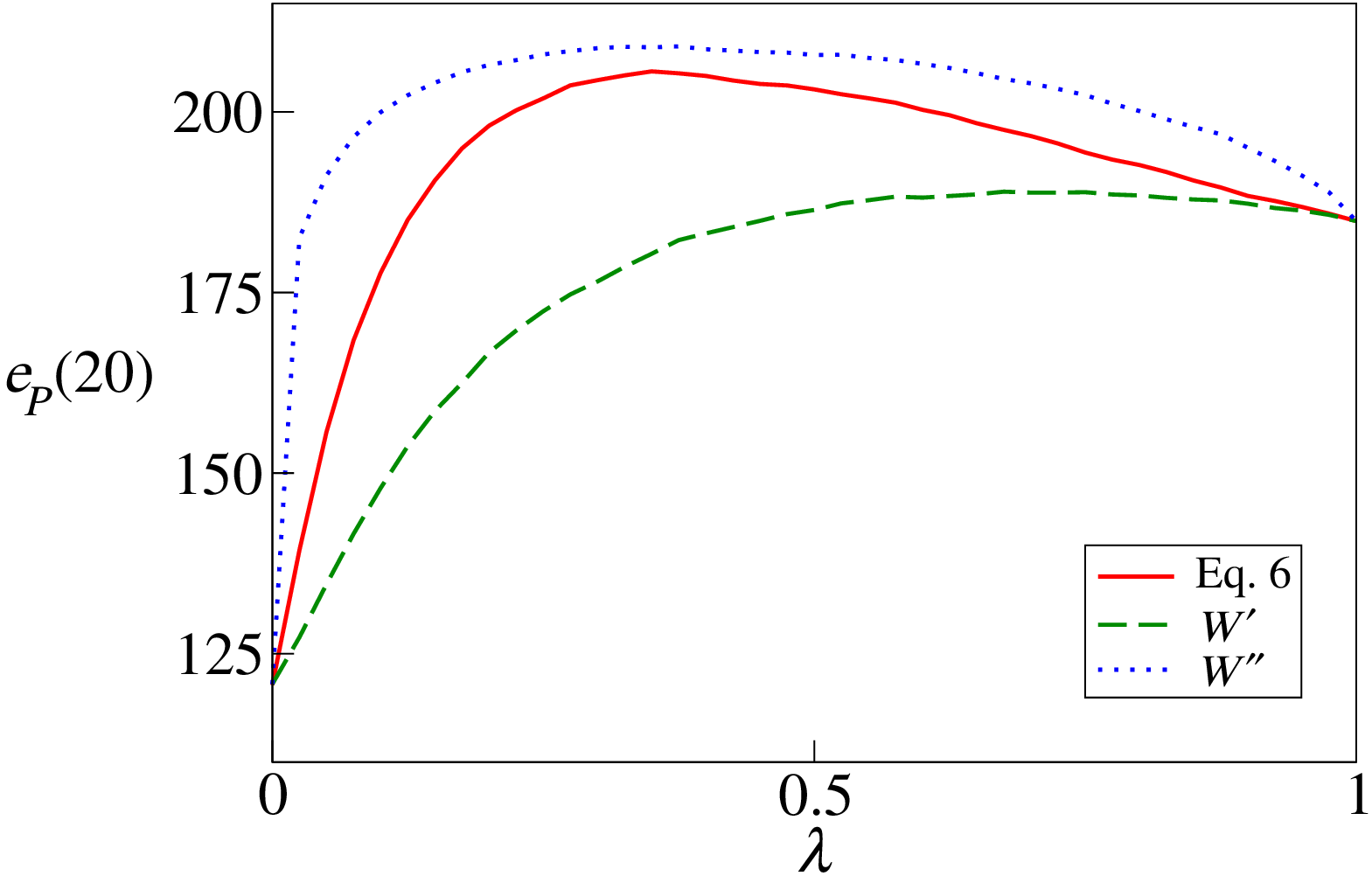}\\[6pt]
\onepic{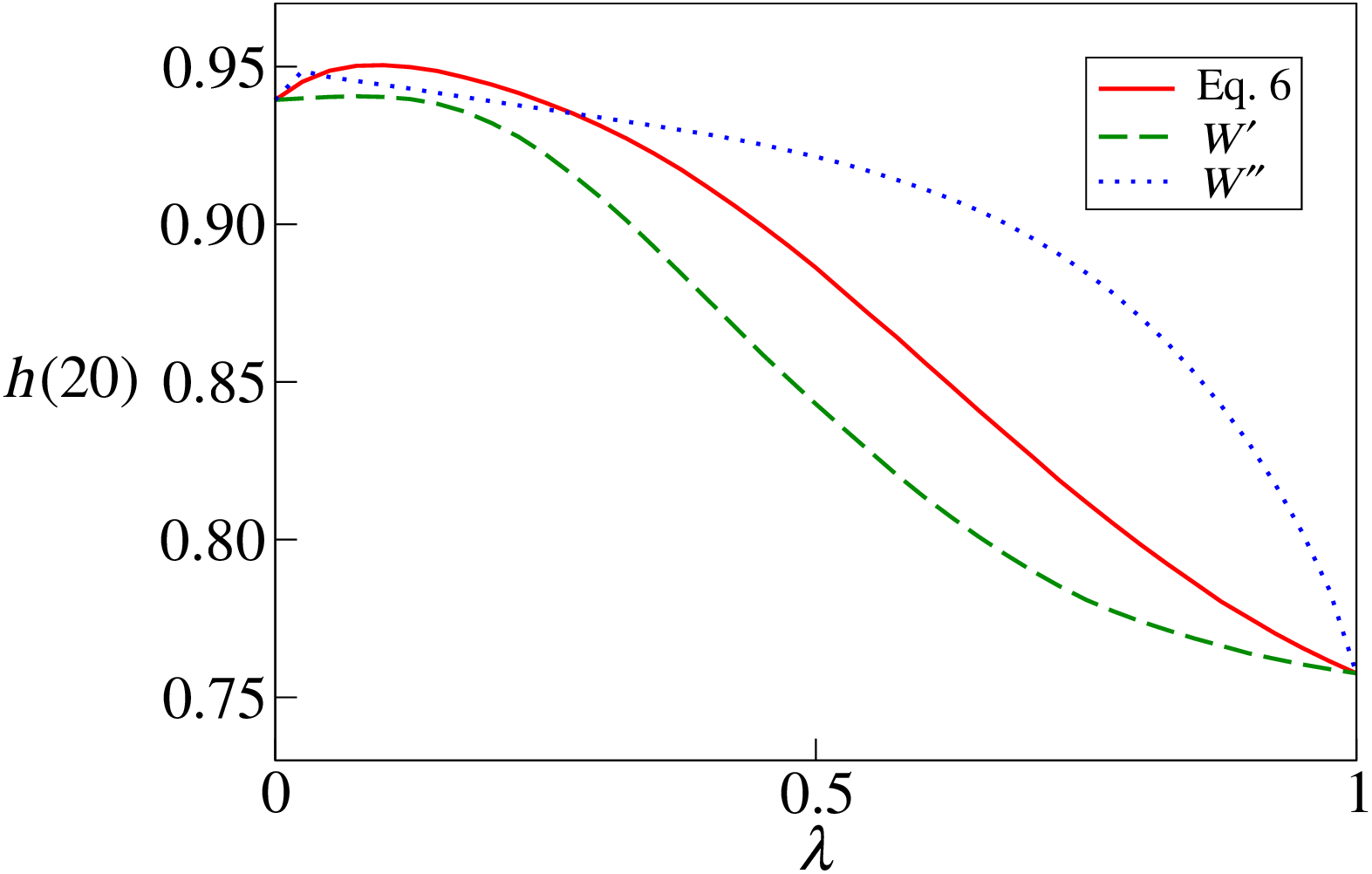} &
\onepic{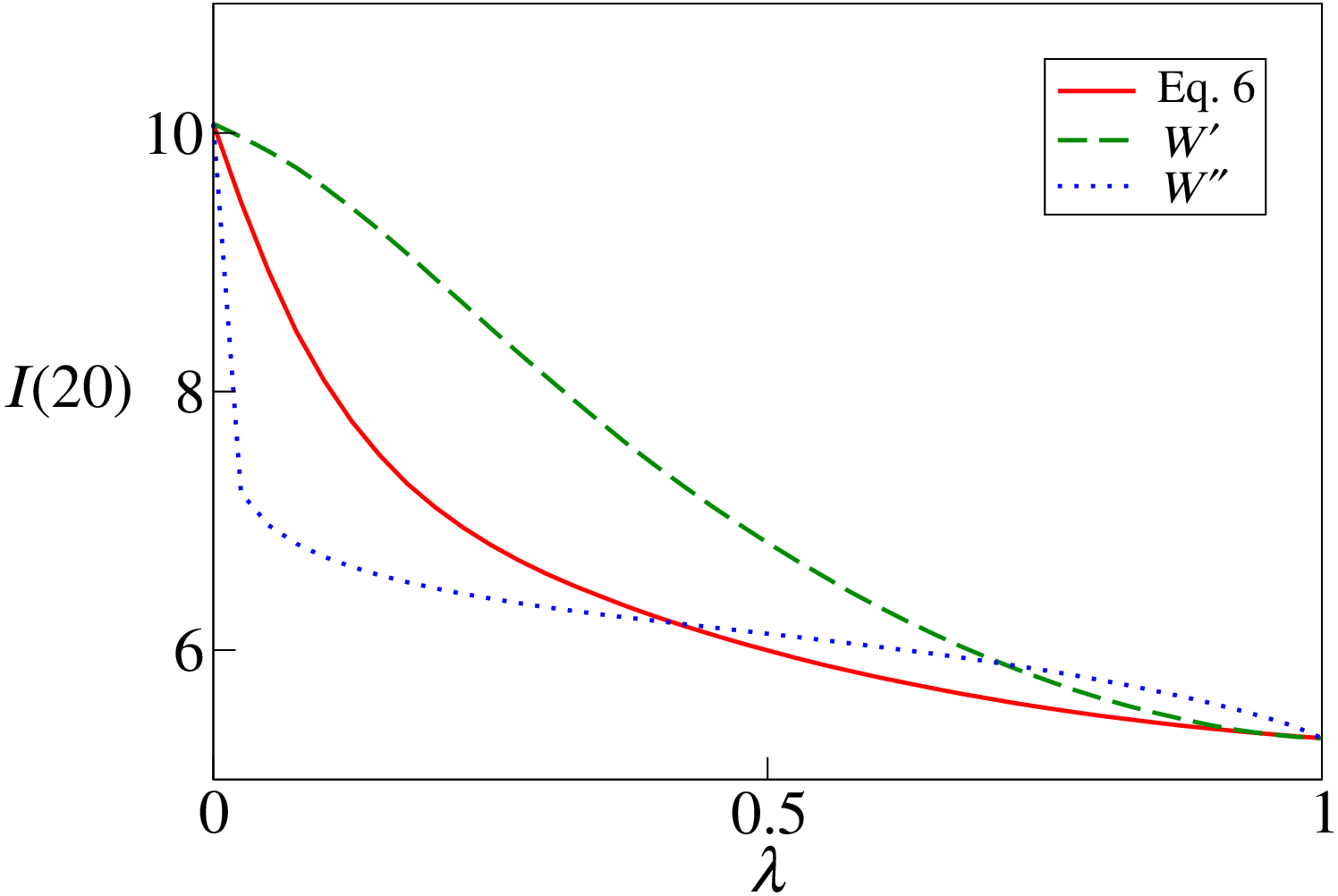}
\end{tabular}
\end{center}
Figure S1. Elegant hybrids of the HeatS and ProbS algorithms
can be created in several ways besides that given in Eq.~6 of
the paper: for example
$W_{\alpha\beta}'=\big(
\frac{1-\lambda}{k_{\alpha}}+\frac{\lambda}{k_{\beta}}\big)
\sum_{j=1}^u a_{\alpha j}a_{\beta j}/k_i$,
or
$W_{\alpha\beta}''=
\frac1{(1-\lambda)k_{\alpha}+\lambda k_{\beta}}
\sum_{j=1}^u a_{\alpha j}a_{\beta j}/k_j$.
While $W_{\alpha\beta}'$ performs well only with respect to
$I(20)$, Eq.~6 and $W_{\alpha\beta}''$ both have their
advantages. However, Eq.~6 is somewhat easier to tune to
different requirements since it varies more slowly and smoothly
with $\lambda$.  The results shown here are for the
RateYourMusic dataset.

\vfill

\begin{center}
\onepic{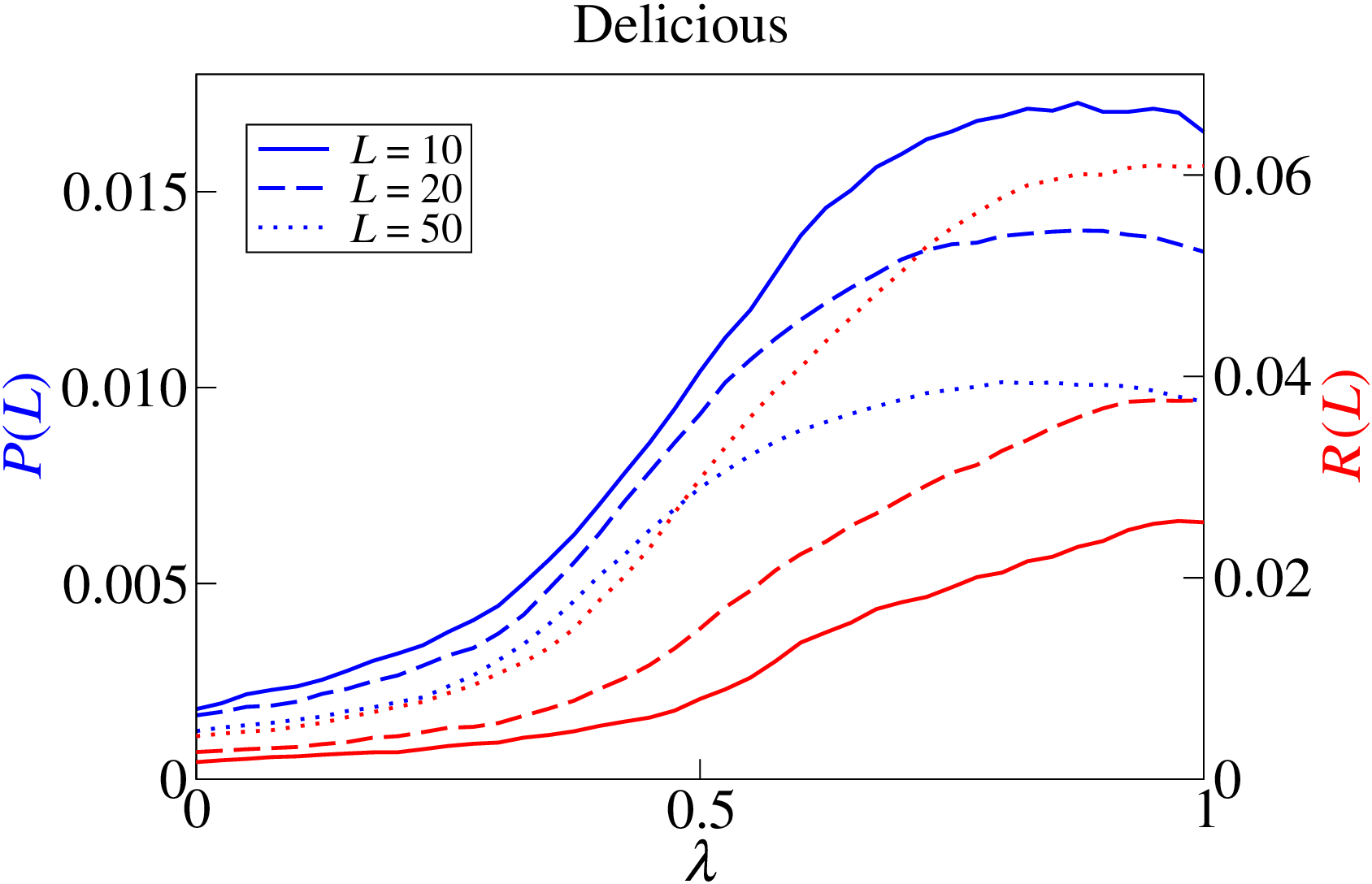} \quad
\onepic{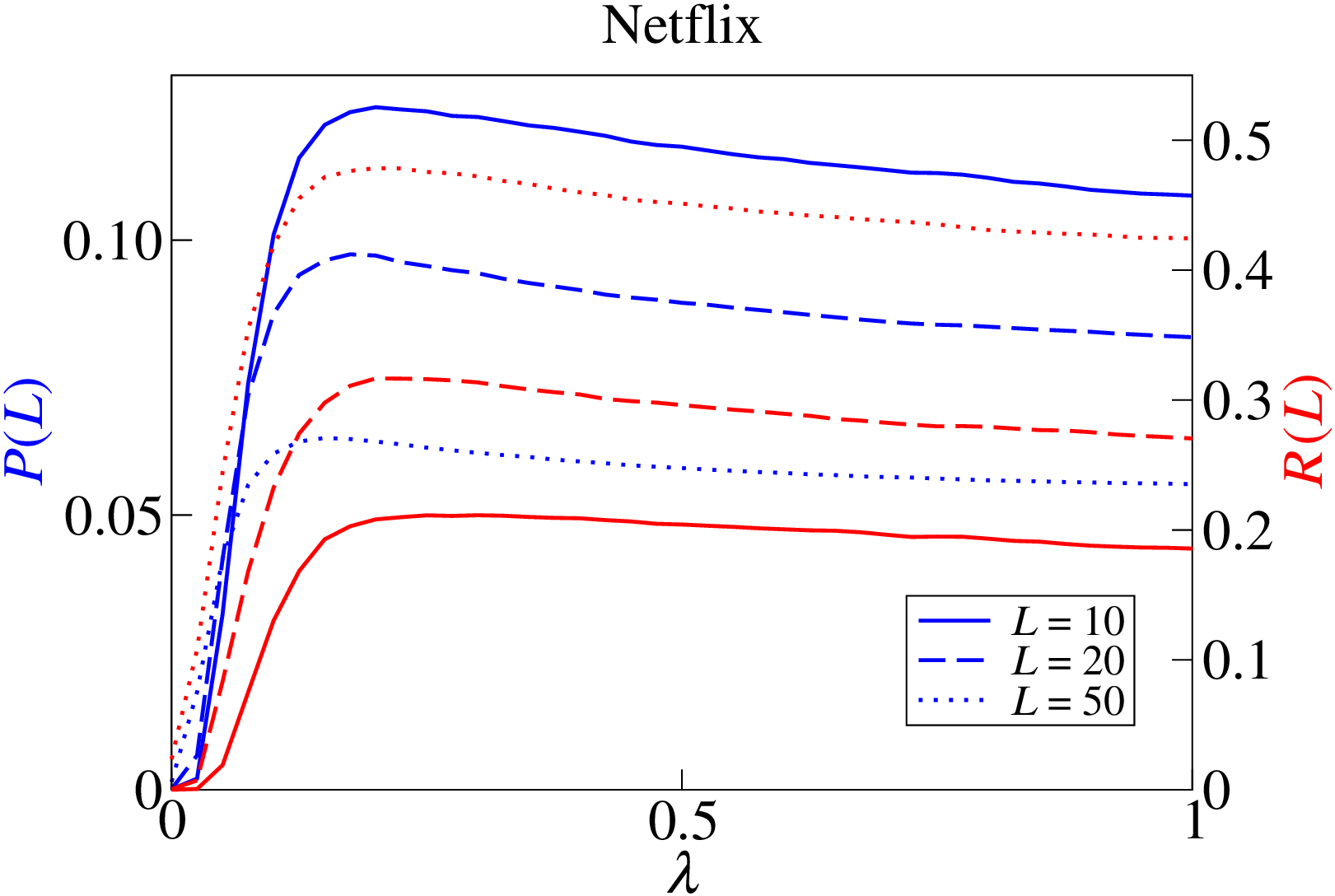}
\end{center}
Figure S2. Precision $P(L)$ and recall $R(L)$ provide
complementary but contrasting measures of accuracy: the former
considers what proportion of selected objects (in our case,
objects in the top $L$ places of the recommendation list) are
relevant, the latter measures what proportion of relevant objects
(deleted links) are selected.  Consequently, recall (red) grows
with $L$, whereas precision (blue) decreases. Here we compare
precision and recall for the HeatS+ProbS hybrid algorithm on the
Delicious and Netflix datasets.  While quantitatively different,
the qualitative performance is very similar for both measures.

\pagebreak

\begin{center}
\onepic{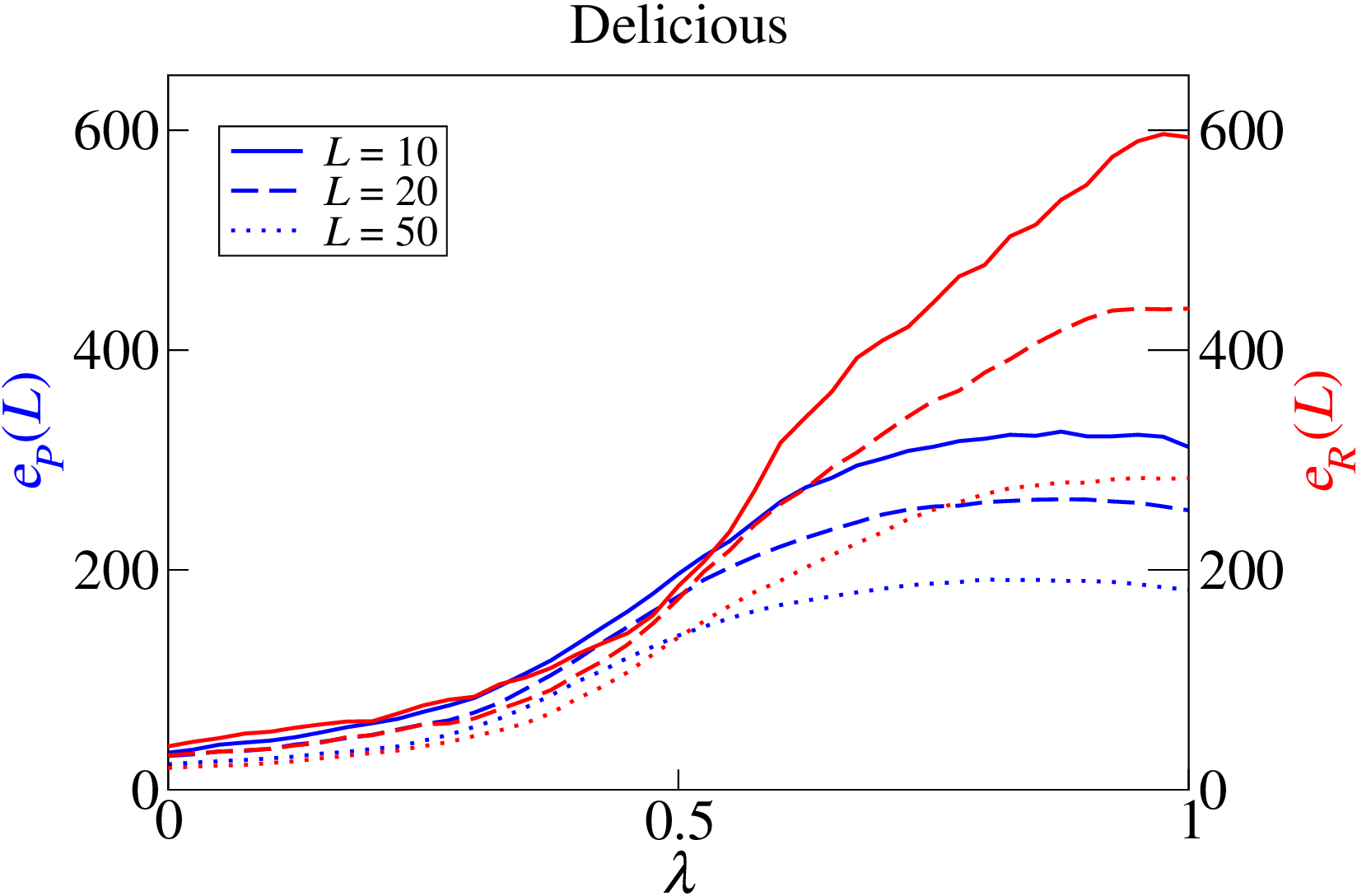} \quad
\onepic{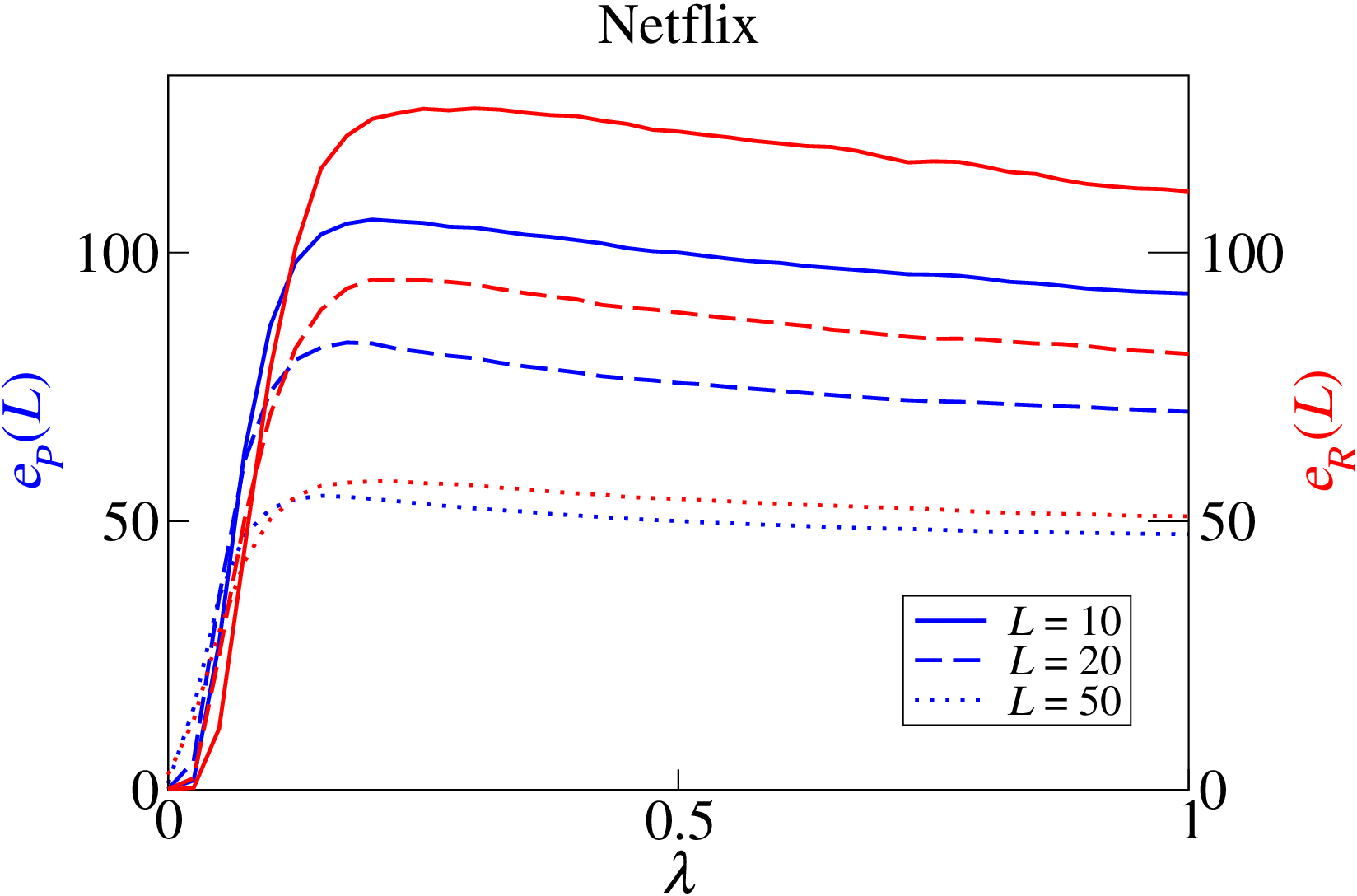}
\end{center}
Figure S3. A more elegant comparison can be obtained by
considering precision and recall \emph{enhancement}, that is,
their values relative to that of randomly-sorted
recommendations: $e_{P}(L) = P(L) \cdot ou/D$ and $e_{R}(L) =
R(L) \cdot o/L$ (Eqs.\ 7a, b in the paper).  Again, qualitative
performance is close, and both of these measures decrease with
increasing $L$, reflecting the inherent difficulty of improving
either measure given a long recommendation list.

\vfill

\begin{center}
\onepic{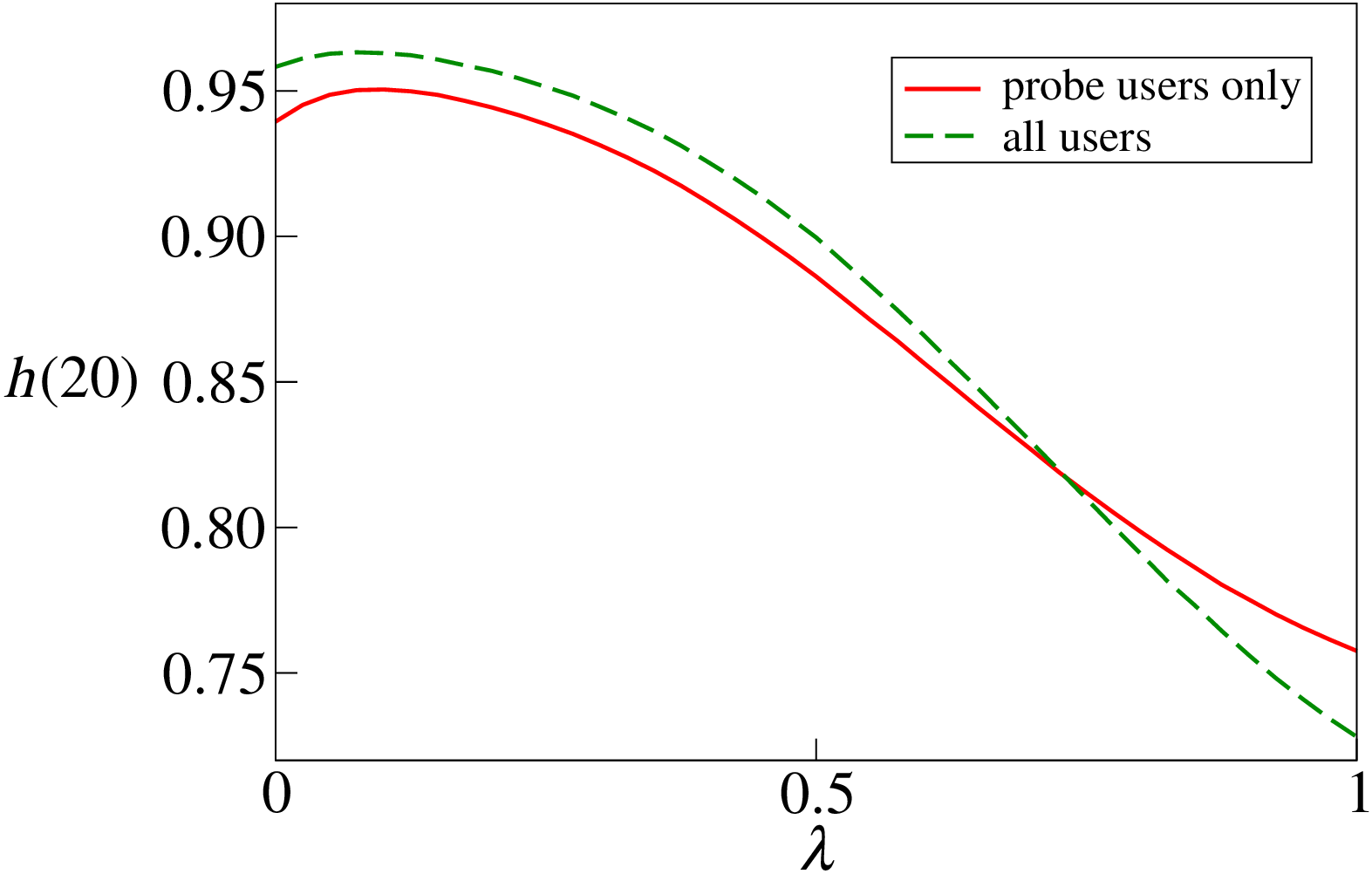}\qquad
\onepic{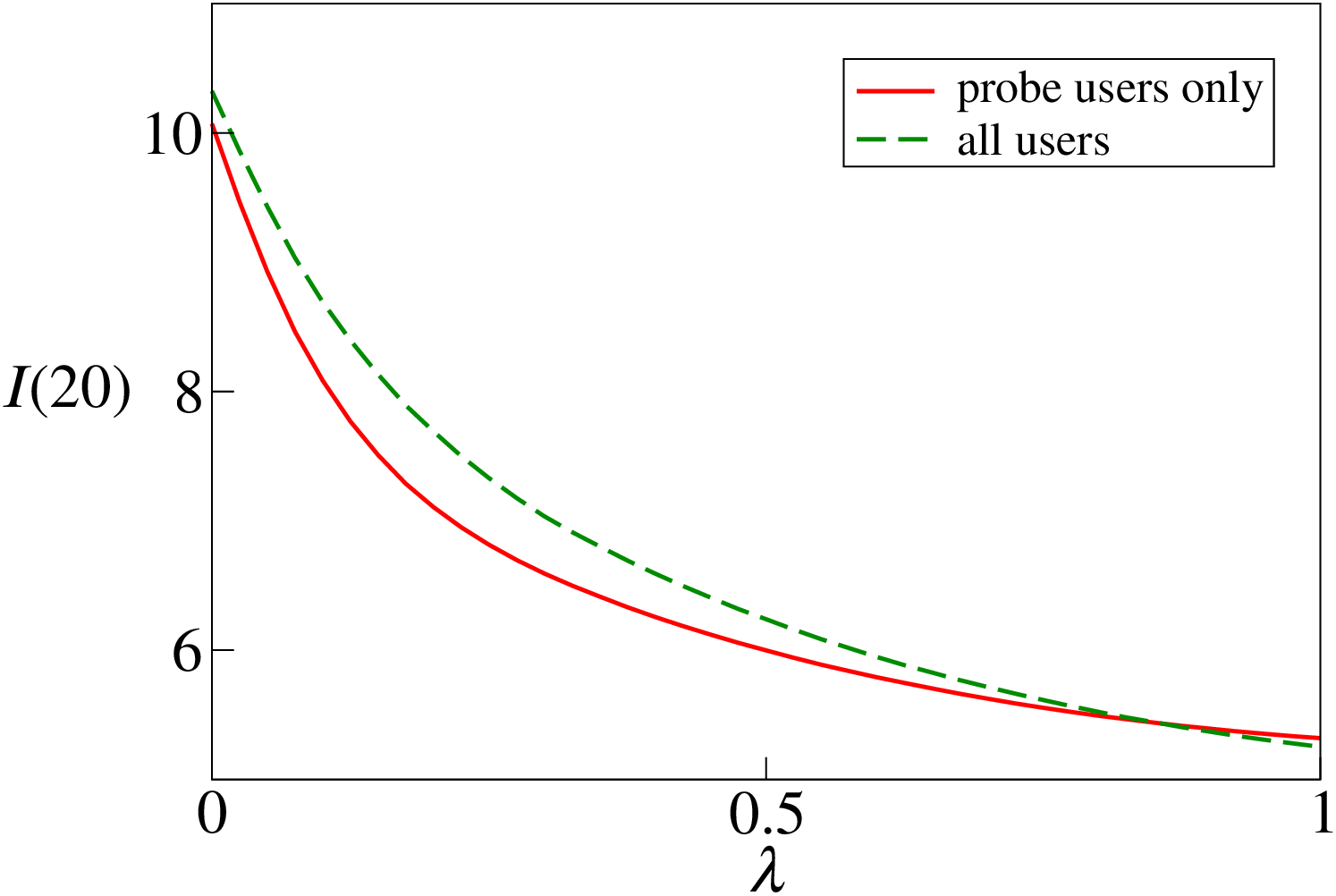}
\end{center}
Figure S4. Comparison of the diversity-related metrics $h(20)$
and $I(20)$ when two different averaging procedures are used:
averaging only over users with at least one deleted link (as
displayed in the paper) and averaging over all users. The
different procedures do not alter the results qualitatively and
make little quantitative difference.  The results shown are for
the RateYourMusic dataset.

\vfill

\begin{center}
\includegraphics[scale=0.55]{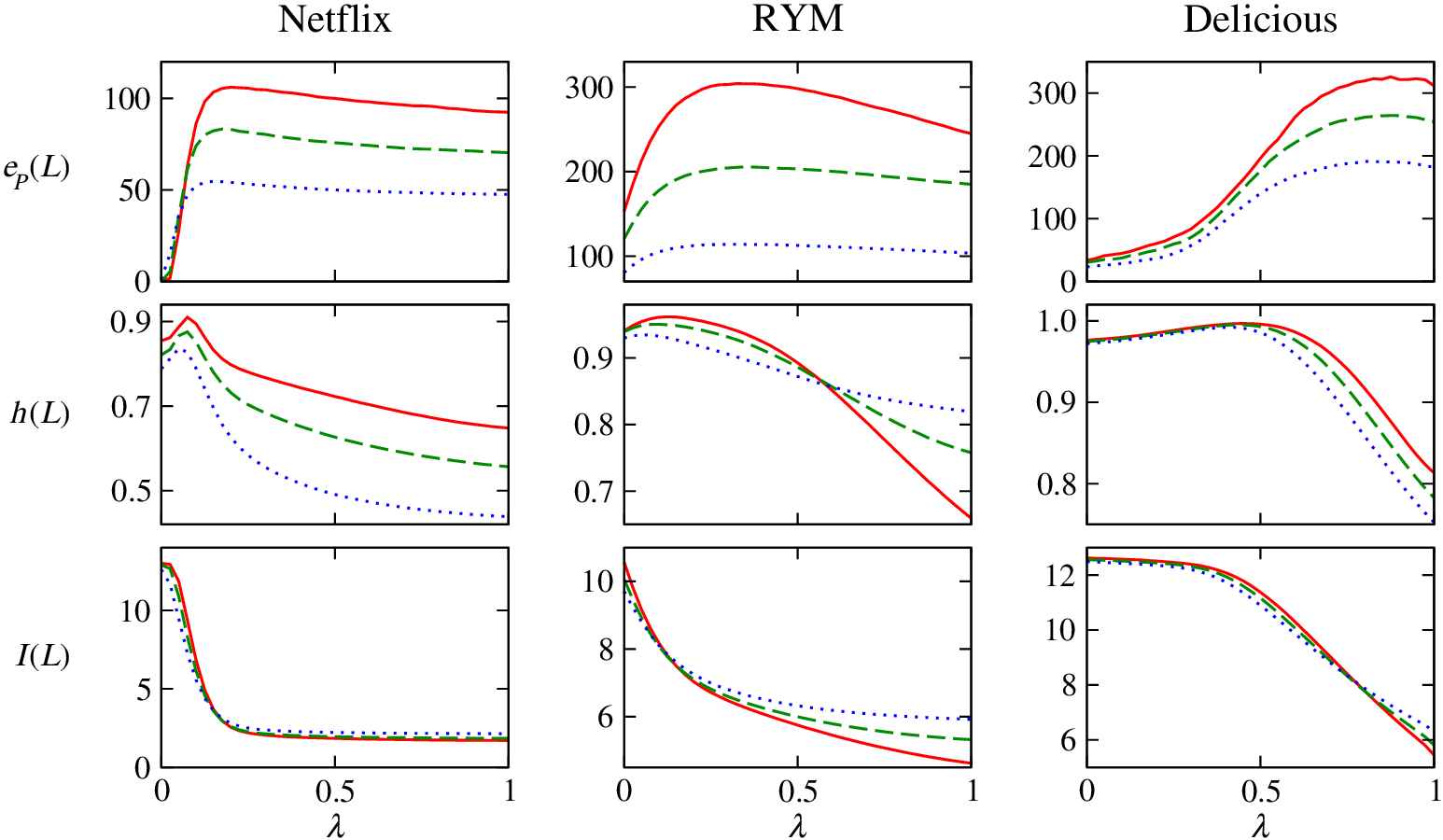}
\end{center}
Figure S5. Comparison of performance metrics for different
lengths $L$ of recommendation lists: $L=10$ (red), $L=20$
(green) and $L=50$ (blue). Strong quantitative differences are
observed for precision enhancement $e_{P}(L)$ and
personalization $h(L)$, but their qualitative behaviour remains
unchanged.  Much smaller differences are observed for surprisal
$I(L)$.

\pagebreak

\begin{center}
\begin{tabular}{rr}
\onepic{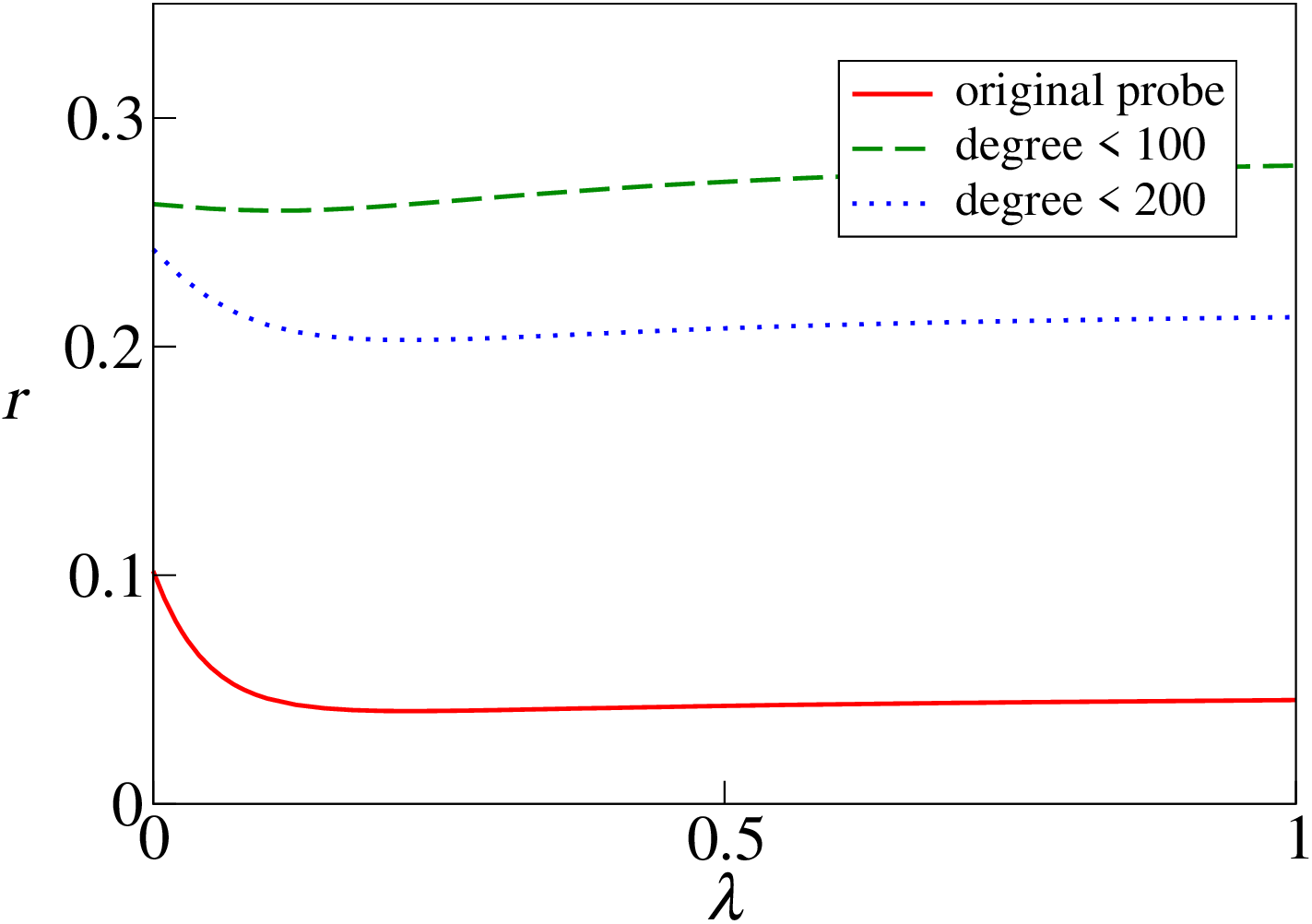} &
\onepic{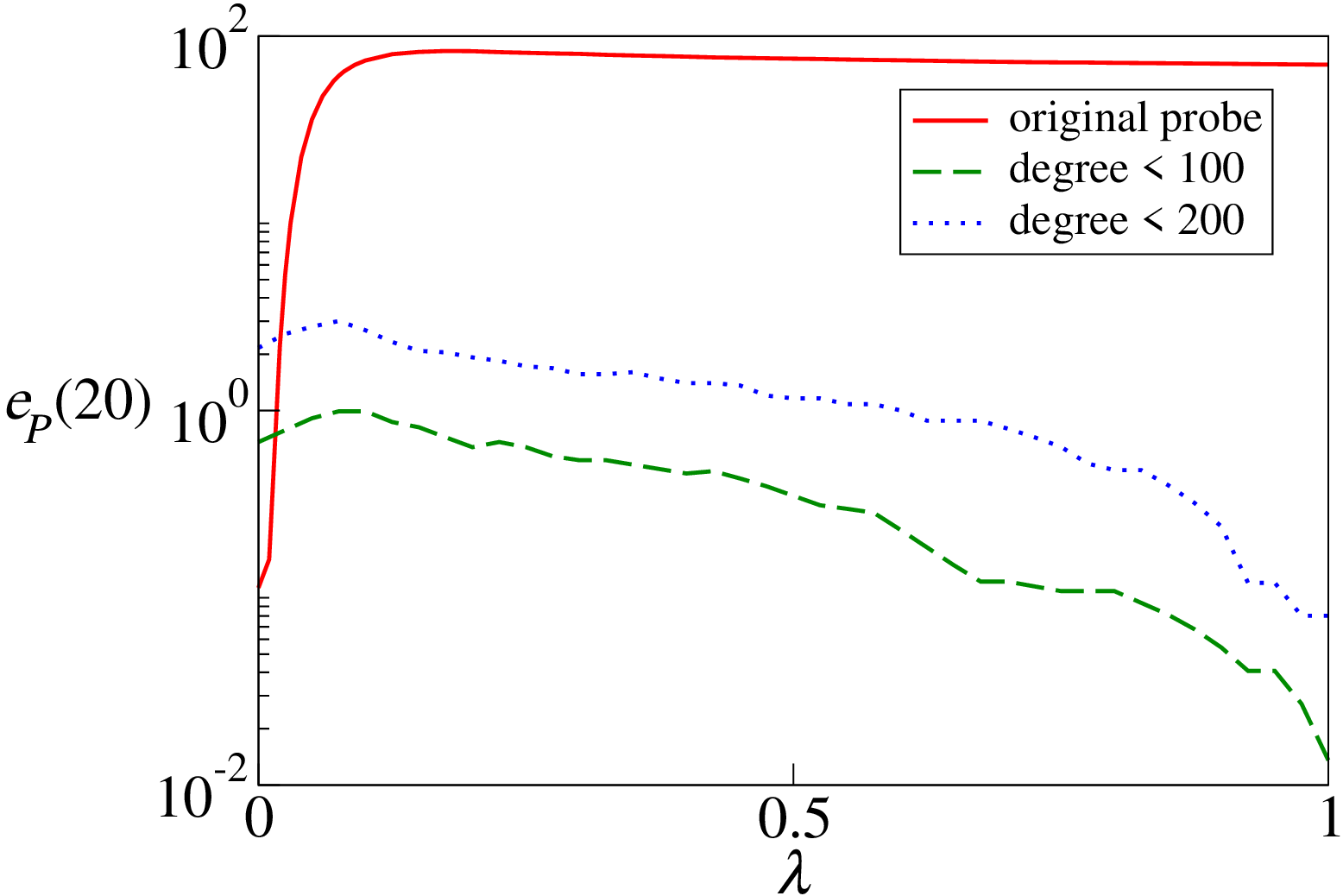}\\
\onepic{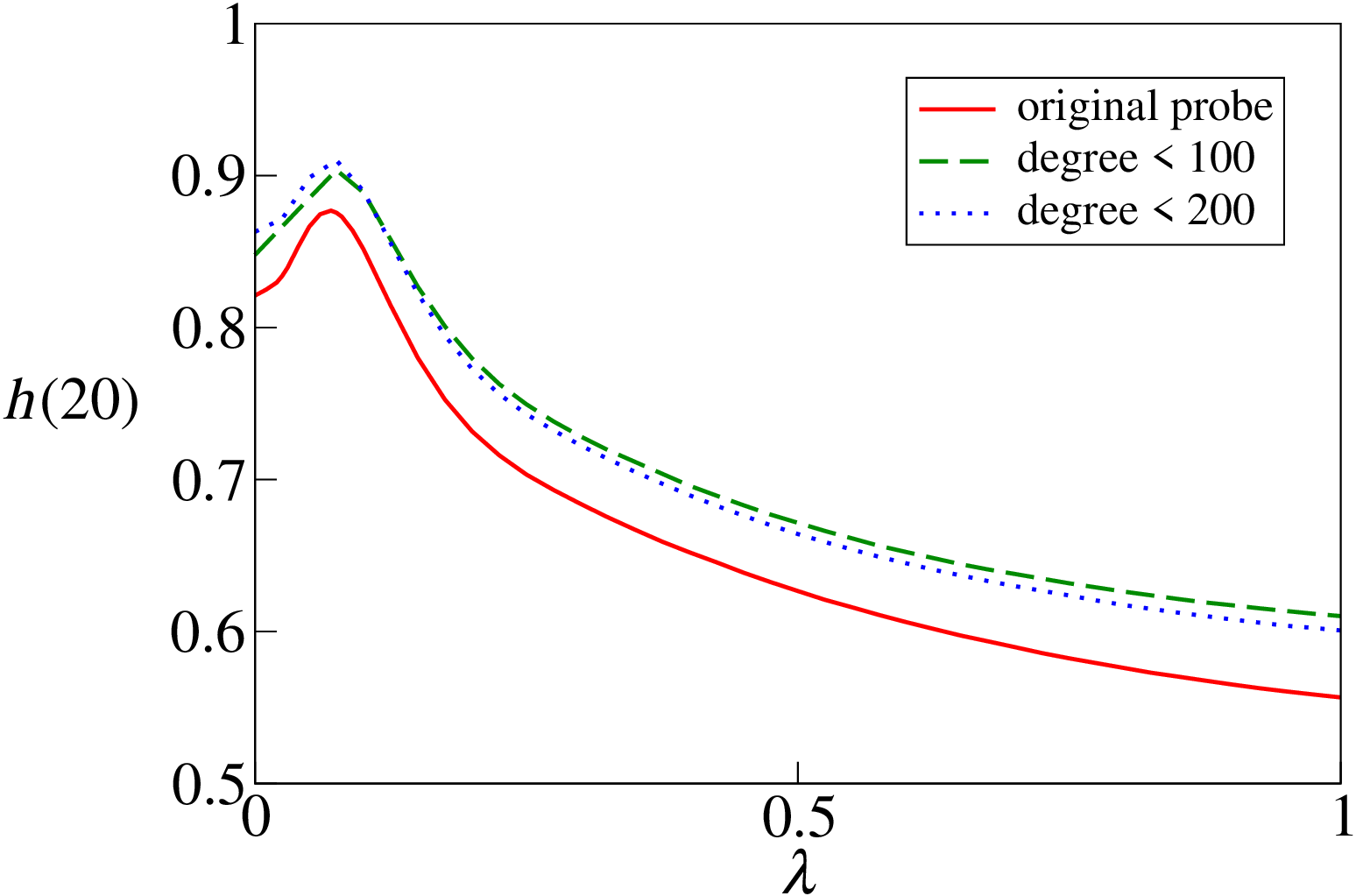} &
\onepic{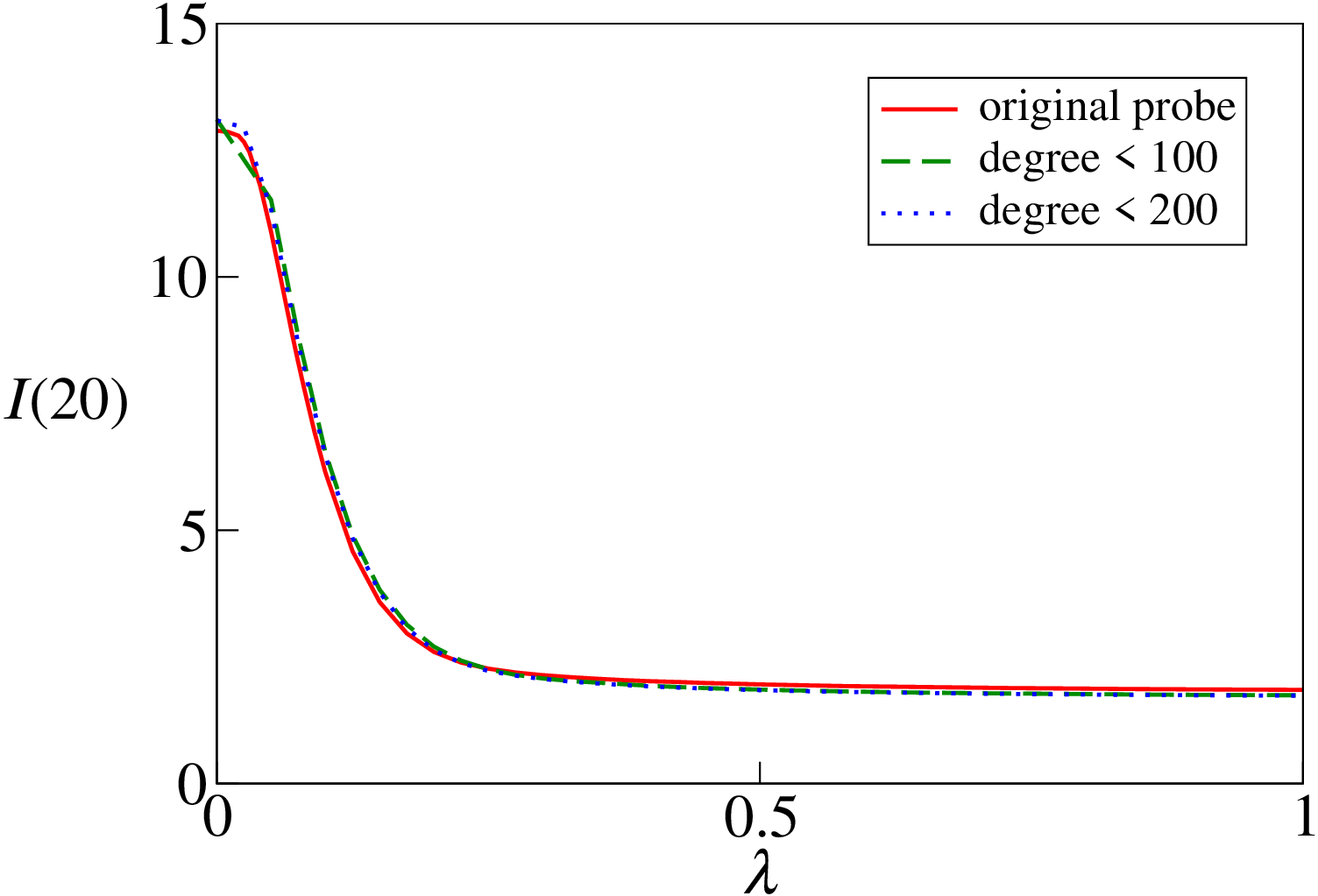}
\end{tabular}
\end{center}
Figure S6. Our accuracy-based metrics all measure in one way or
another the recovery of links deleted from the dataset. Purely
random deletion will inevitably favor high-degree (popular)
objects, with their greater proportion of links, and
consequently methods that favor popular items will appear
to provide higher accuracy. To study this effect, we created two
special probe sets consisting of links only to objects whose
degree was less than some threshold (either 100 or 200): links
to these objects were deleted with probability $0.5$, while
links to higher-degree objects were left untouched. The result
is a general decrease in accuracy for all
algorithms---unsurprisingly, since rarer links are inherently
harder to recover---but also a reversal of performance, with the
low-degree-favoring HeatS now providing much higher accuracy
than the high-degree-oriented ProbS, USim and GRank.
The results shown here are for the Netflix dataset.

\vfill

\begin{center}
\begin{tabular}{rr}
\onepic{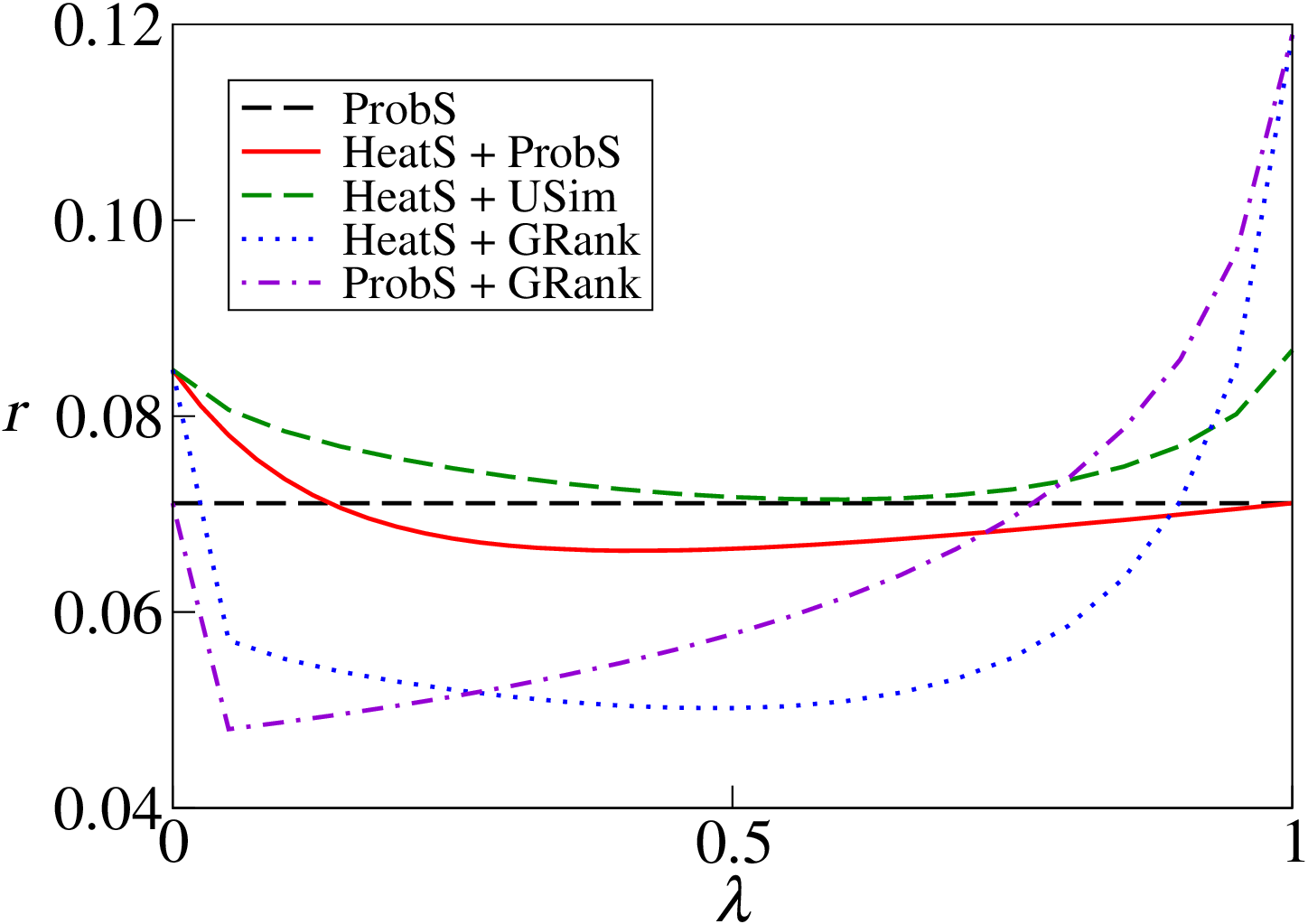} &
\onepic{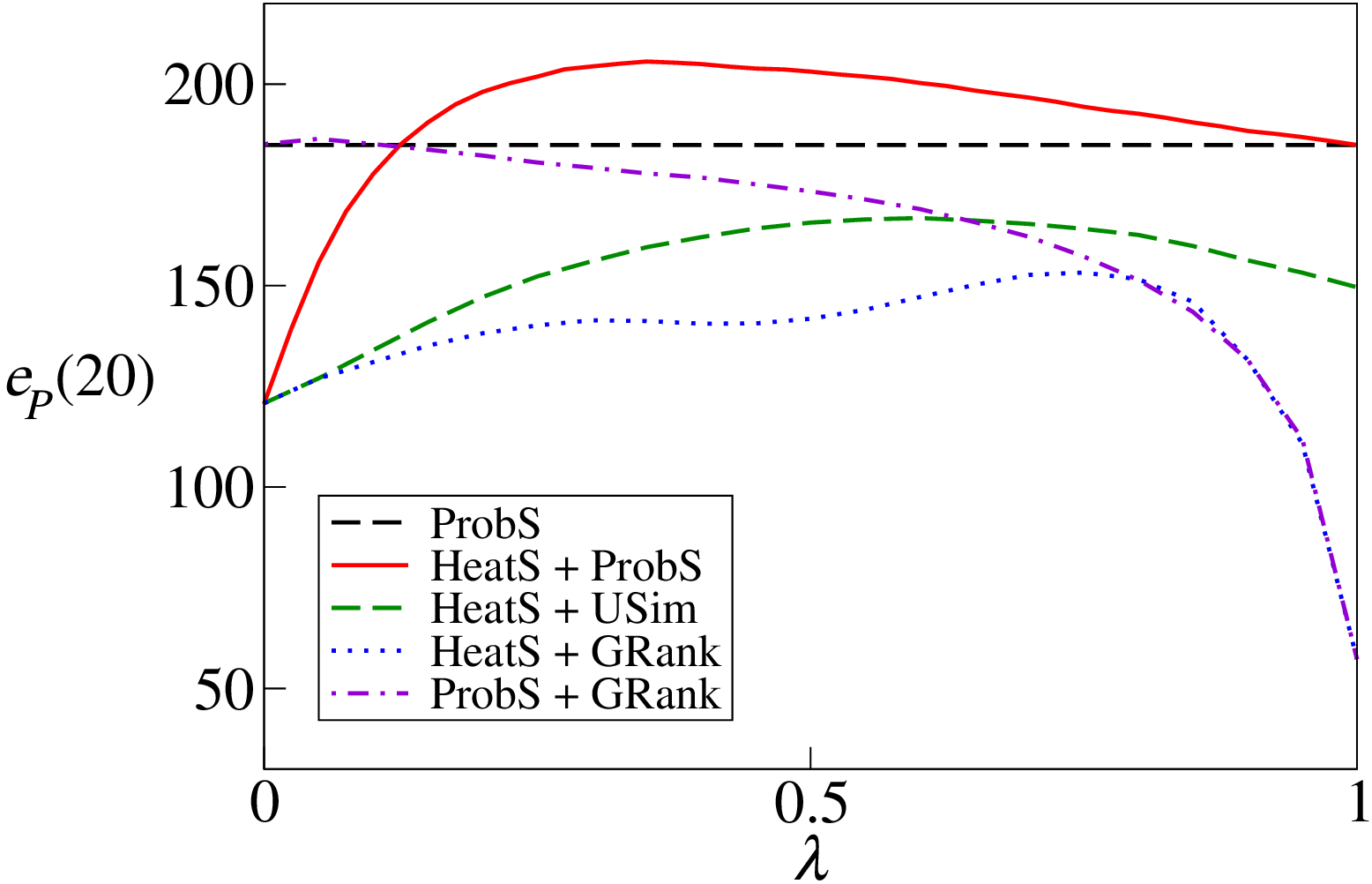}\\
\onepic{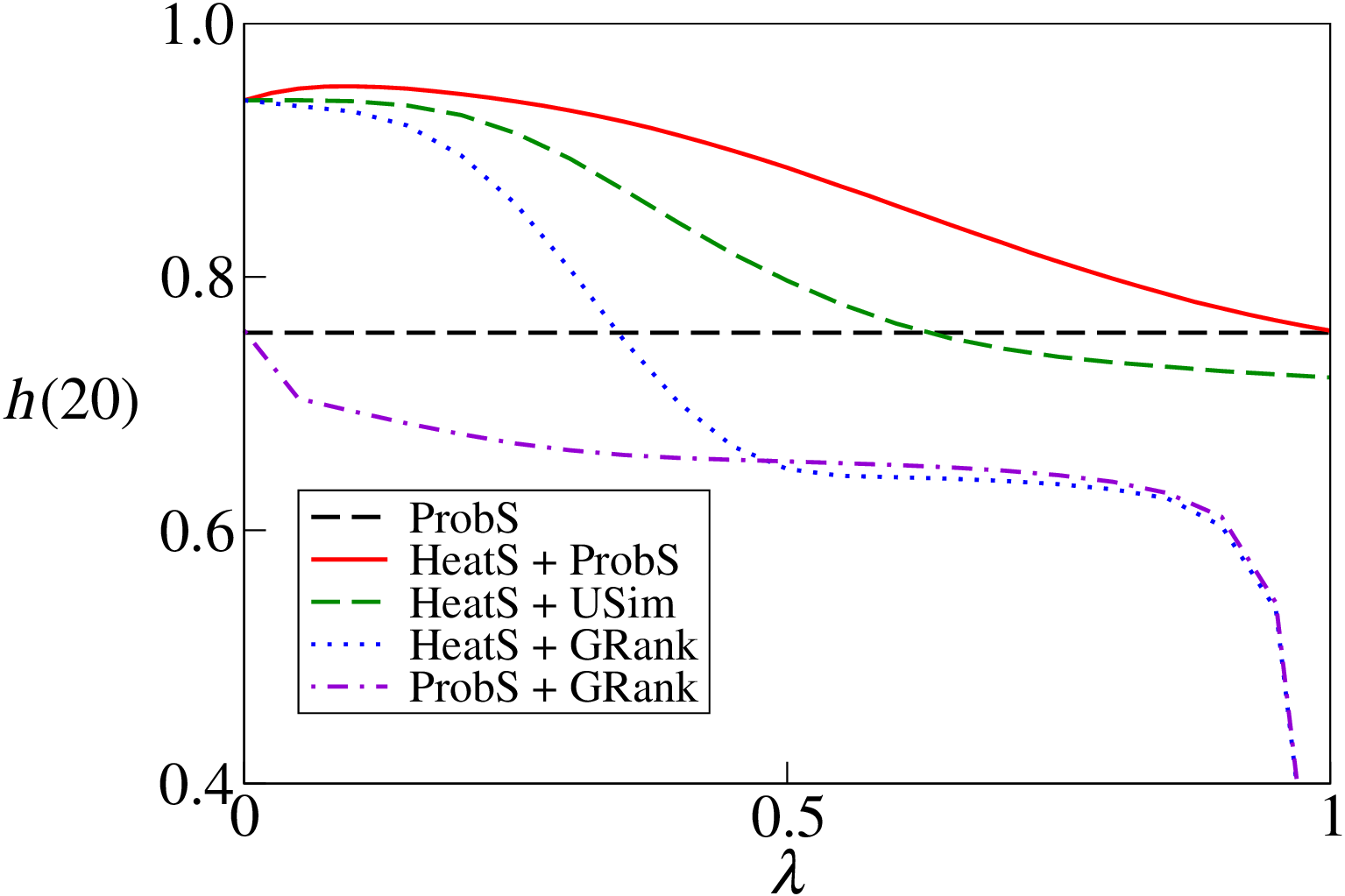} &
\onepic{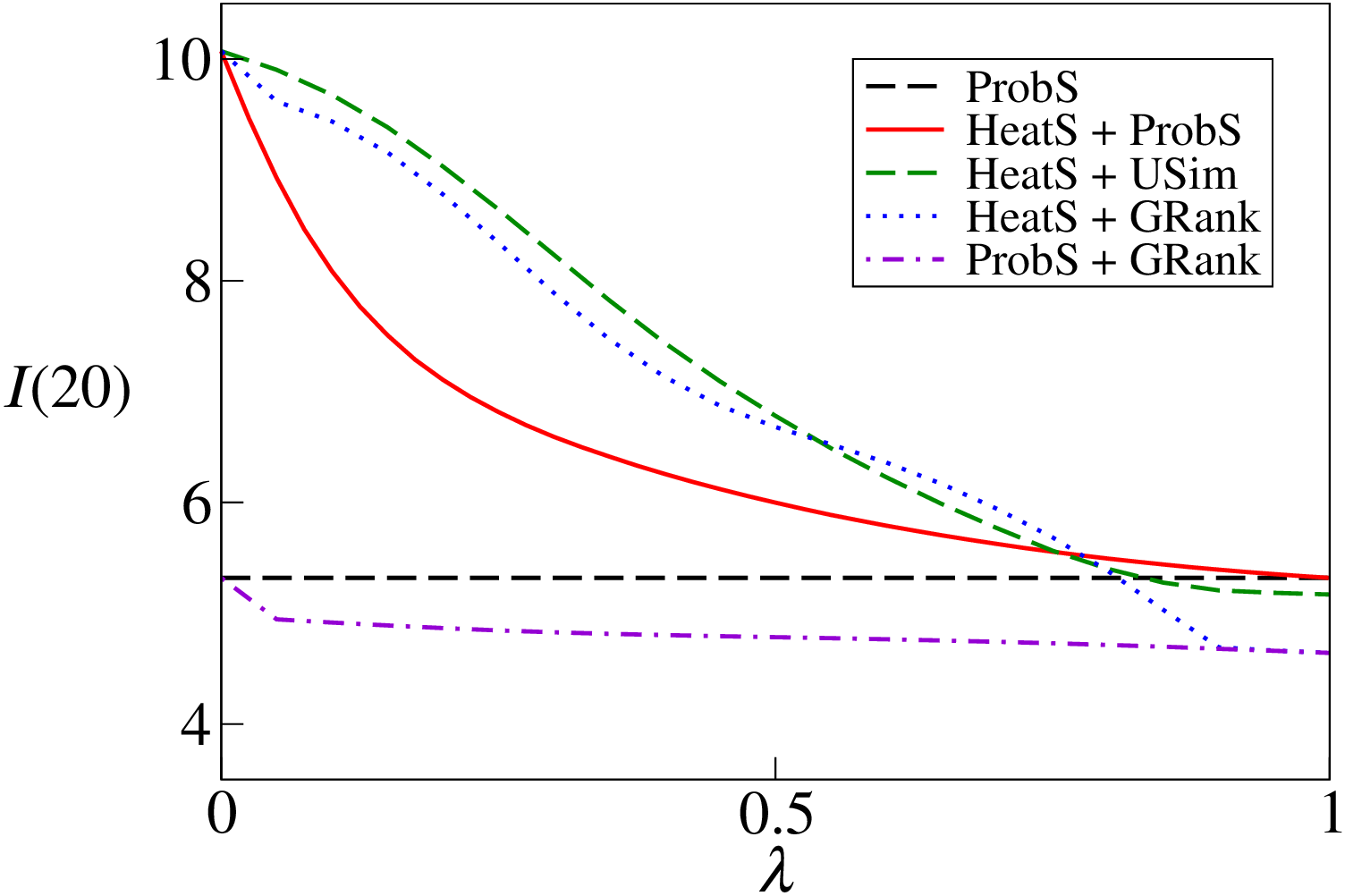}
\end{tabular}
\end{center}
Figure S7. In addition to HeatS+ProbS, various other hybrids
were created and tested using the method of Eq.~5 in the
paper, where for hybrid X+Y, $\lambda=0$ corresponds to  pure X
and $\lambda=1$ pure Y. The results shown here are for the
Netflix dataset. The HeatS+USim hybrid offers similar but weaker
performance compared to HeatS+ProbS; combinations of GRank with
other methods produce significant improvements in $r$, the
recovery of deleted links, but show little or no improvement of
precision enhancement $e_{P}(L)$ and poor results in
diversity-related metrics. We can conclude that the proposed
HeatS+ProbS hybrid is not only computationally convenient but
also performs better than combinations of the other methods
studied.

\end{document}